\documentclass[manuscript]{aastex631}

\usepackage{graphicx}

\definecolor{darkred}{RGB}{175,0,0}

\shorttitle{EUV Resonant Excitation During Flares}
\shortauthors{Seaton et al.}
\graphicspath{{./}{figures/}}

\begin{document}

\title{Evidence of Extreme Ultraviolet Resonant Excitation in the Middle Corona During A Solar Flare}

\correspondingauthor{Daniel B. Seaton}

\author[0000-0002-0494-2025]{Daniel B. Seaton}
\affiliation{Southwest Research Institute, Boulder, CO 80302, USA}

\author[0000-0003-1759-4354]{Cooper Downs}
\affiliation{Predictive Science Inc., San Diego, CA 92121, USA}

\author[0000-0002-4125-0204]{Giulio Del Zanna}
\affiliation{University of Cambridge, Cambridge, UK}
\affiliation{School of Physics \& Astronomy, University of Leicester, Leicester  LE1 7RH, UK}

\author[0000-0002-0631-2393]{Matthew J. West}
\affiliation{Southwest Research Institute, Boulder, CO 80302, USA}
\affiliation{European Space Agency, European Space Research and Technology Centre, Noordwijk, Netherlands}

\author[0000-0002-5305-9466]{Edward M. B. Thiemann}
\affiliation{Laboratory for Atmospheric and Space Physics, University of Colorado, Boulder, CO, 80305, USA}

\author[0000-0001-8702-8273]{Amir Caspi}
\affiliation{Southwest Research Institute, Boulder, CO 80302, USA}

\author[0000-0001-7416-2895]{Edward E. DeLuca}
\affiliation{Center for Astrophysics $\vert$ Harvard \& Smithsonian, 60 Garden Street, Cambridge, MA, USA}

\author[0000-0001-9638-3082]{Leon Golub}
\affiliation{Center for Astrophysics $\vert$ Harvard \& Smithsonian, 60 Garden Street, Cambridge, MA, USA}

\author[0000-0002-3783-5509]{James Paul Mason}
\affiliation{Johns Hopkins University, Applied Physics Laboratory, Laurel, MD}

\author[0000-0001-8504-2725]{Ritesh Patel}
\affiliation{Southwest Research Institute, Boulder, CO 80302, USA}

\author[0000-0002-6903-6832]{Katharine K. Reeves}
\affiliation{Center for Astrophysics $\vert$ Harvard \& Smithsonian, 60 Garden Street, Cambridge, MA, USA}

\author[0000-0002-8748-2123]{Yeimy J. Rivera}
\affiliation{Center for Astrophysics $\vert$ Harvard \& Smithsonian, 60 Garden Street, Cambridge, MA, USA}

\author[0000-0002-6172-0517]{Sabrina Savage}
\affiliation{NASA Marshall Space Flight Center, Huntsville, AL, 35808, USA}



\begin{abstract}
We present observations of an eruptive solar flare on 2016~January~6 that occurred behind the solar limb from the perspective of the Earth, but was well observed by \textit{STEREO} and the Solar Extreme Ultraviolet Monitor on the \textit{Mars Atmosphere and {Volatile} EvolutioN (MAVEN)} mission. Light curves showing the evolution of the flare's irradiance as a function of time taken by MAVEN are well correlated with the brightness evolution of fan structures observed in the PROBA2 SWAP 174~\AA\ passband, suggesting that the radiance of structures near the flare site was influenced by emission from the flare. Because SWAP did not directly observe the flare itself, this event represents a rare opportunity to study the evolution of emission near a flare without the risk of instrumental scattered light contaminating the observations. We analyze this evolution and implement a simple model to explore the possibility that resonant excitation (or resonant scattering) plays an important role in driving coronal EUV emission during flaring events. Our modeling shows that for a large flare, resonant excitation could increase emission from nearby structures by about 45\%, consistent with our findings that the involved structures observed by SWAP increased in brightness by about 60\% during the flare. We conclude that resonant excitation may play an important role in driving coronal EUV emission under certain circumstances and should be accounted for in models and emission-based analysis tools.
\end{abstract}


\section{Introduction}\label{sec:intro}
Despite decades of availability of extreme-ultraviolet (EUV) observations of the solar corona, questions about the nature of emission from the corona -- particularly in the \textit{middle corona} \citep{West2023} -- remain unanswered. It is known that emission due to collisional excitation dominates in the low corona \citep{delzanna2018a}, but the significance of other emission mechanisms is poorly constrained. This question is particularly important for tools that produce temperature diagnostics in the middle corona, {between about 1.5 and 6~R$_\sun$}, which historically have generally assumed that all EUV emission arises from collisional excitation\footnote{{Throughout this paper we adopt the definitions of \citet{West2023} for inner, middle, and outer coronae, with the inner corona defined as below 1.5~R$_\sun$, and the outer corona beginning at 6~R$_\sun$.}}. If resonant excitation (RE), also known as resonant scattering, or other mechanisms account for a significant fraction of the emission in some regions of the corona, our diagnostic tools must be revised.  This also adds complexity to the modeling and analysis of coronal EUV emission, as collisionally-dominated emission is proportional simply to density squared, while generally the intensity of a spectral line depends on a fine balance between the local density, the photo-exciting (PE) radiance, plus other factors such as the local velocity. Such complexities are well known in the UV where they have been used to enhance the diagnostic potential of observations, e.g., measuring the outflow plasma velocity via Doppler dimming \citep[e.g.,][]{Strachan1993}. RE is even more important for the visible light line emission \citep[see, e.g.,][]{Boe2023} and the near-infrared \citep[e.g.,][]{Judge1998}, but has generally not been considered or fully exploited for its diagnostic potential in the EUV.

Although a few studies using the Transition Region and Coronal Explorer \citep[TRACE;][]{handy1999} suggested that RE may contribute up to half of all EUV coronal emission \citep{schrijver_mcmullen_2000}, most authors have argued that contributions from mechanisms other than collisional excitation are negligible in the low corona \citep[e.g.,][]{deforest2009, delzanna2018b}. However, recent studies that examined observed \citep{goryaev2014, Seaton2021, Auchere2023} and modeled \citep{Gilly2020} EUV emission in the middle corona suggest that RE plays an important role with increasing height in the corona.

The ratios of these collisional to RE contributions have only been inferred observationally using scaling laws \citep[see][]{Seaton2021} as it is not simple to disentangle their relative importance directly. One approach that ought to provide a more direct way to disentangle these contributions is to observe the effect of a bright EUV flare on the emission from structures that are not involved in the flare itself \citep[e.g.,][]{goryaev2018}. If the radiance from such structures varies in direct proportion to the radiance from the flare itself, the most likely explanation is that this is due to RE of flare-produced photons.

Unfortunately, the design of most EUV telescopes introduces a significant complication to making such measurements: some light is scattered into the very broad wings of the instrumental point-spread function (PSF) both by the microroughness of the mirrors in the telescope \citep[for additional discussion see, e.g.,][]{seaton2013, martinez-galarce2010} and by diffraction of the filter support grids \citep{schwartz2014}. A bright flare can scatter a significant amount of light inside the instrument, leading to confusion between resonantly scattered flare emission and instrumental stray light. \citep[For additional discussion of the effects of scattered light on observations of the extended corona, see][]{Auchere2023}.

Even the use of PSF deconvolution methods, which could remove much of this stray light, may not help resolve this confusion. These methods rely on having accurate knowledge of signal across the entire image plane. Bright flares often lead to saturation of the detector near the core of the flare, and thus are not compatible with this requirement, making it nearly impossible to accurately remove scattered light from flare images.

However, in very rare cases there is a way around this difficulty! In this paper we present observations of a bright EUV flare on 2016~January~6, which occurred behind the solar limb from the Earth's perspective. As a result, no light originating directly from the flare itself entered the optics of the Sun Watcher with Active Pixels and Image Processing \citep[SWAP;][]{SeatonSwap2013, Halain2013} onboard the \textit{Project for On-Board Autonomy 2 (PROBA2)} spacecraft. This serendipitous event provides an opportunity to observe the effects of flare-driven RE without any instrumental complications.

During the flare, the Solar Extreme Ultraviolet Monitor \citep{Eparvier2015} on NASA's Mars Atmosphere and Volatile EvolutioN (MAVEN) mission had a clear view of the flare. This allowed us to straightforwardly track the relationship between the flare irradiance and brightening in the middle corona observed by SWAP, even though SWAP itself could not observe the flare.

\section{Observations}\label{sec:obs}
\begin{figure}
    \centering
    \begin{interactive}{animation}{XRT_BEM20160106103306_175502.mpg}
    \includegraphics[width=0.9\textwidth]{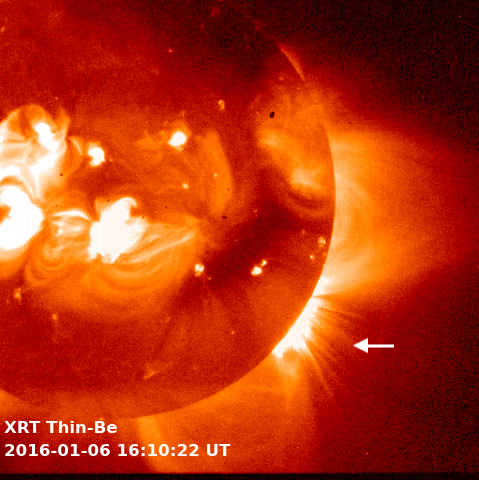}
    \end{interactive}
    \caption{{Observations of SADs in XRT (indicated by a white arrow in the static version of this figure), above a flare that cannot be observed from Earth. Such features are often associated with strong flares, but in this case the flare occurred completely behind the limb from the Earth's perspective.}}
    \label{fig:XRT_Flare}
\end{figure}

On 2016~January~6 the X-ray Telescope \citep[XRT;][]{Golub2007} on \textit{Hinode} made an unusual observation of what appeared to be a very strong eruption at the southwest limb of the Sun, including supra-arcade downflows (SADs; \citealt{Savage2012, Shen2022}), with no accompanying X-ray flare signature (see Figure~\ref{fig:XRT_Flare} and the accompanying animation). In fact, the GOES X-ray flux at this time initially decreased, and then remained largely flat for several hours. (Although the XRT images were dominated by flare-associated dynamics, including supra-arcade downflows, that rendered them unusable for the specific analysis we describe below, it was the visibility of the eruption in these observations that permitted us to identify this as a candidate event and so we mention them here.) Observations from the Large Angle and Spectrometric Coronagraph \citep[LASCO;][]{Brueckner1995} reveal an associated coronal mass ejection (CME), with speeds in excess of 2000~km/s as reported by the Computer-Aided CME Tracking Tool \citep[CACTus;][]{Robbrecht2009}.

\begin{figure}
    \centering
    \includegraphics[width=0.75\textwidth]{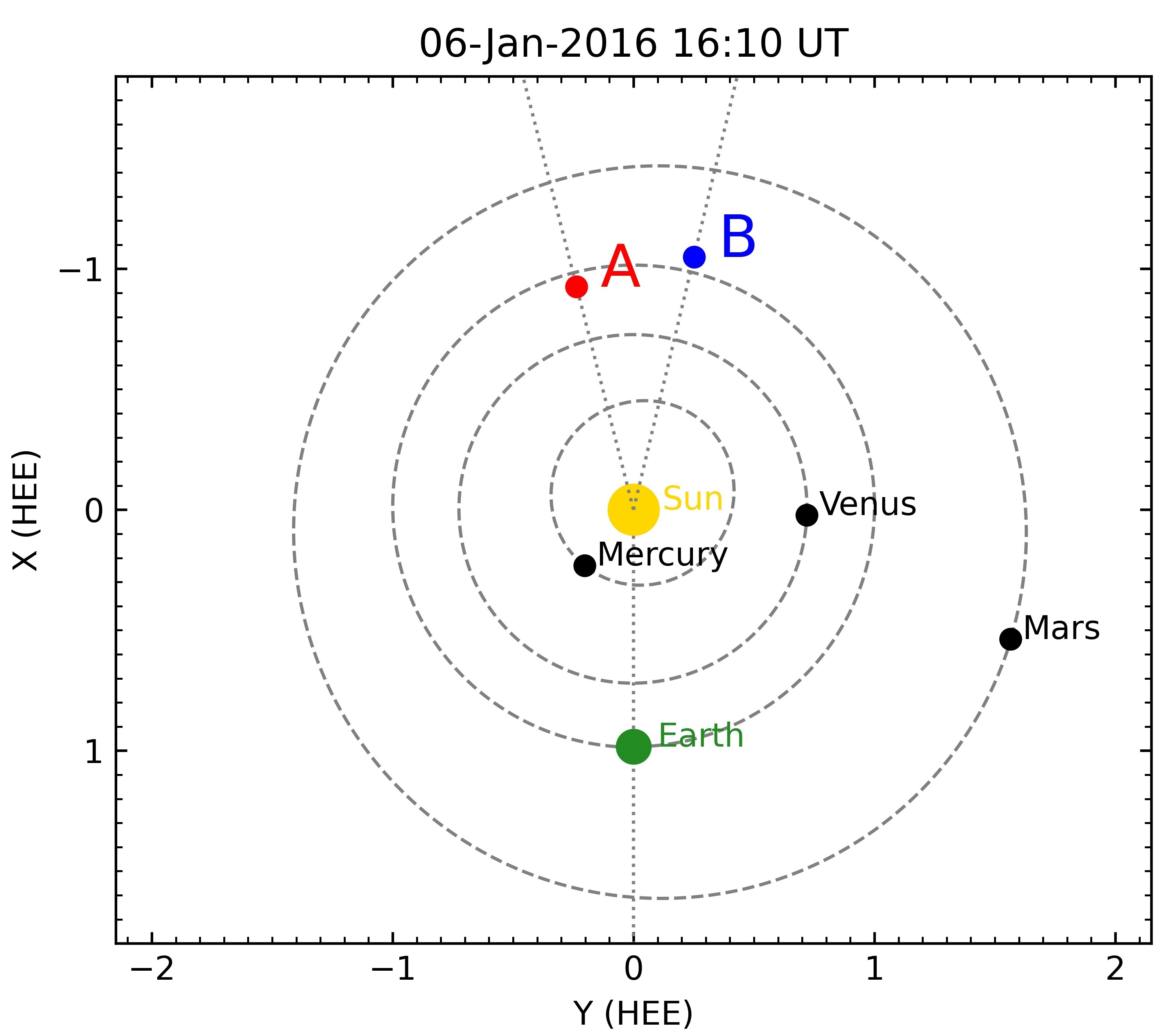}
    \caption{Location of key viewpoints and spacecraft at the time of the flare. STEREO-A is close to the far side of the Sun, while MAVEN, at Mars, has a clear view of the face of the Sun where the flare occurred. PROBA2 is located in Earth orbit. {Note that although the figure shows STEREO-B, that spacecraft has been defunct since October 2014.}}
    \label{fig:Positions}
\end{figure}

\begin{figure}
    \centering
    \includegraphics[width=0.9\textwidth]{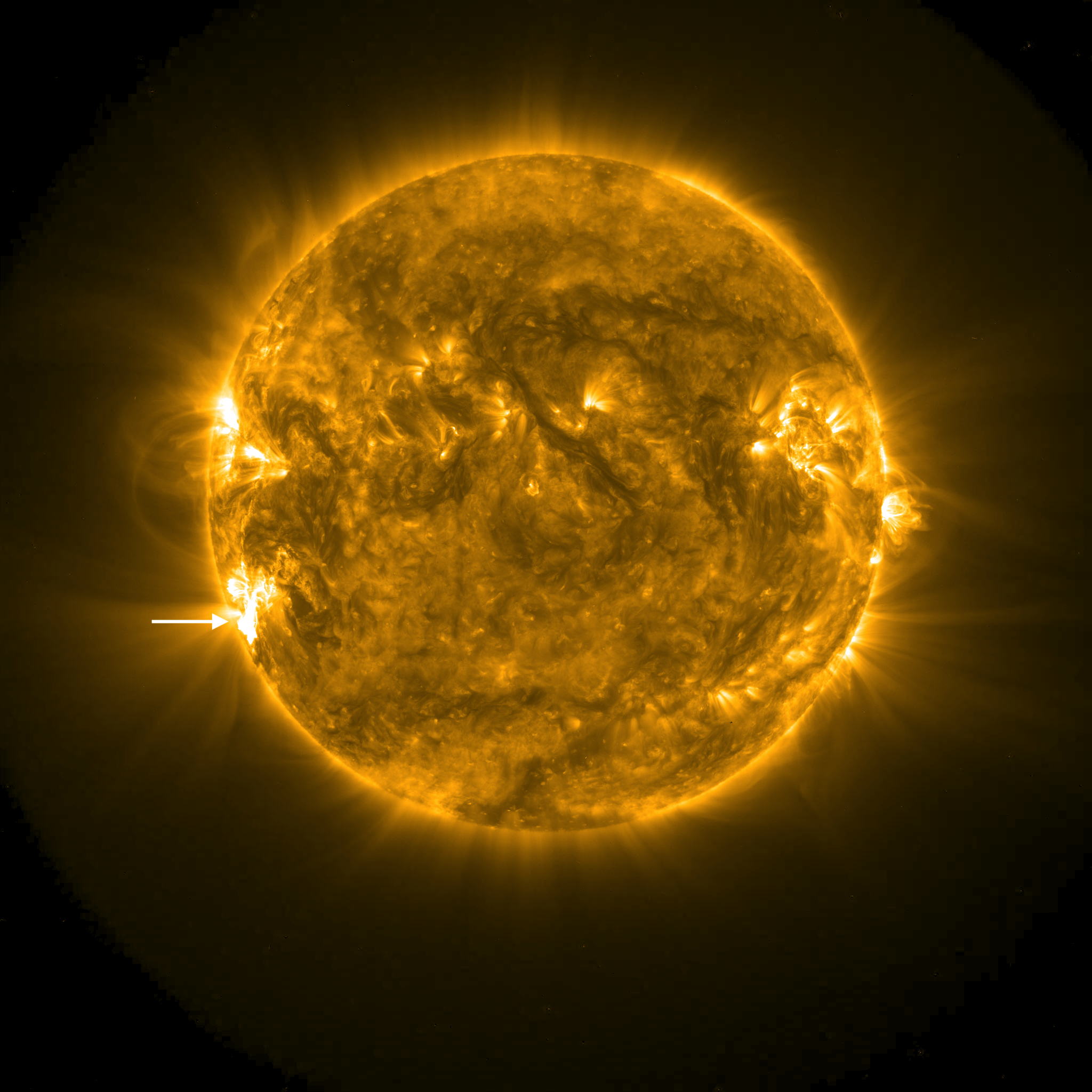}
    \caption{{A STEREO-A/EUVI 171~\AA\ image of the flare at 16:14~UT near the East limb of the Sun (indicated by the arrow in the figure). Figure~\ref{fig:EUVI-195-movie} movie shows the full evolution of the event as seen in STEREO-A's 195~\AA\ passband (171~\AA\ movies of the flare are unavailable due to the low cadence in this channel).}}
    \label{fig:EUVI-171}
\end{figure}

\begin{figure}
\centering
\begin{interactive}{animation}{stereoa_195_20160106.mp4}
    \includegraphics[width=0.9\textwidth]{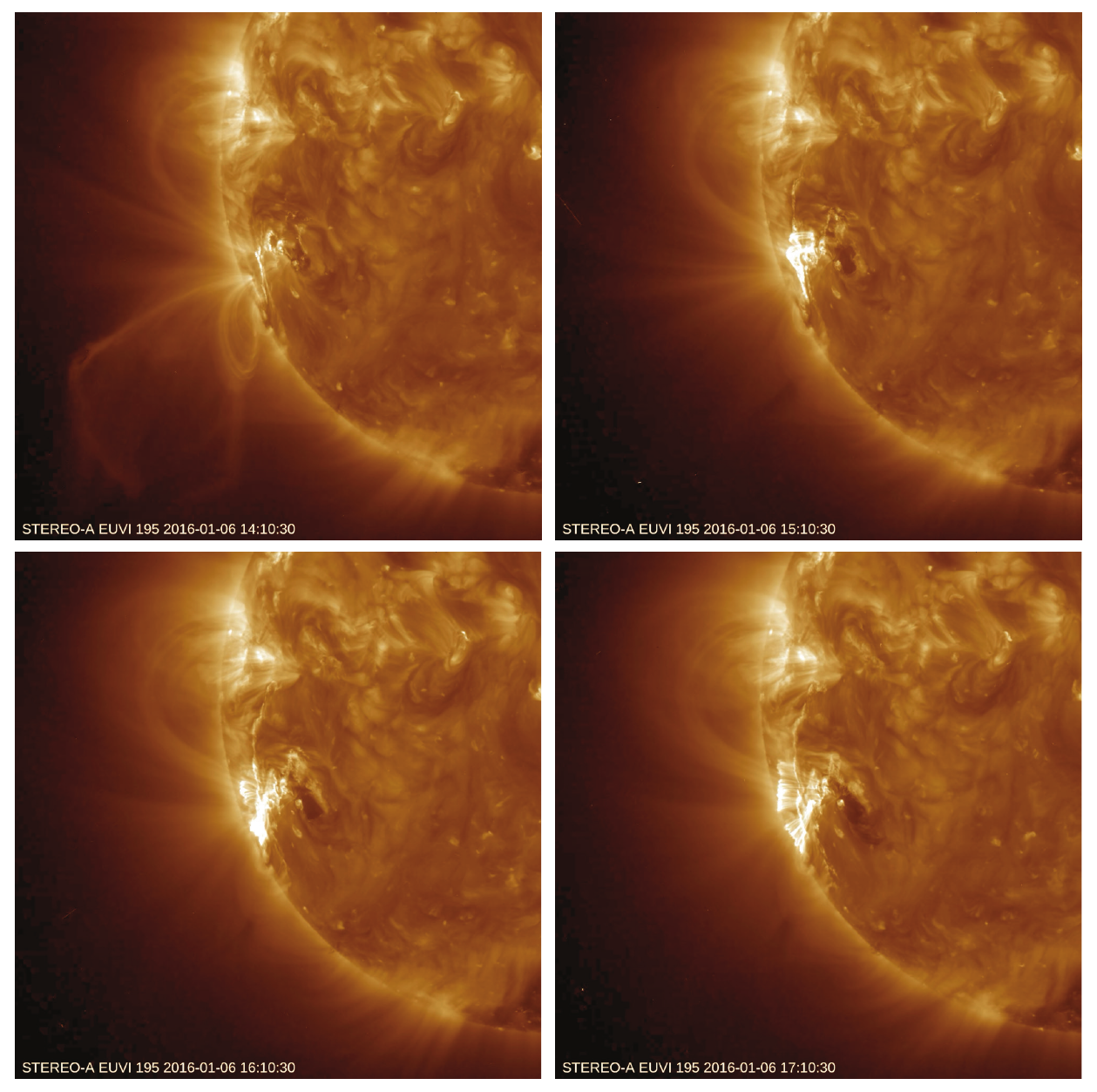}
\end{interactive}
\caption{{STEREO-A/EUVI 195~\AA\ observation of the evolution of the flare, near the east limb of the Sun. The lower right panel corresponds to roughly the same time as the image in Figure~\ref{fig:EUVI-171}.}}
\label{fig:EUVI-195-movie}
\end{figure}

Observations from the Extreme Ultraviolet Imager \citep[EUVI;][]{Howard2008} on the \textit{Solar-Terrestrial Relations Observatory-A (STEREO-A)} mission, located on the far side of the Sun during the event (Figure~\ref{fig:Positions}) reveal the flare's location near, but behind, the limb from the Earth's point of view. Observations in {EUVI's 171~\AA\ (Figure~\ref{fig:EUVI-171}) and 195~\AA\ (Figure~\ref{fig:EUVI-195-movie})} channels reveal the onset of an eruption and growth of a post-eruptive arcade. The 171~\AA\ observations, obtained later in the evolution of the event, clearly reveal a bright fan structure behind the flare site itself, near the East limb, from STEREO-A's perspective.

\begin{figure}
    \centering
    \begin{interactive}{animation}{swap_20160106.mp4}
        \includegraphics[width=0.9\textwidth]{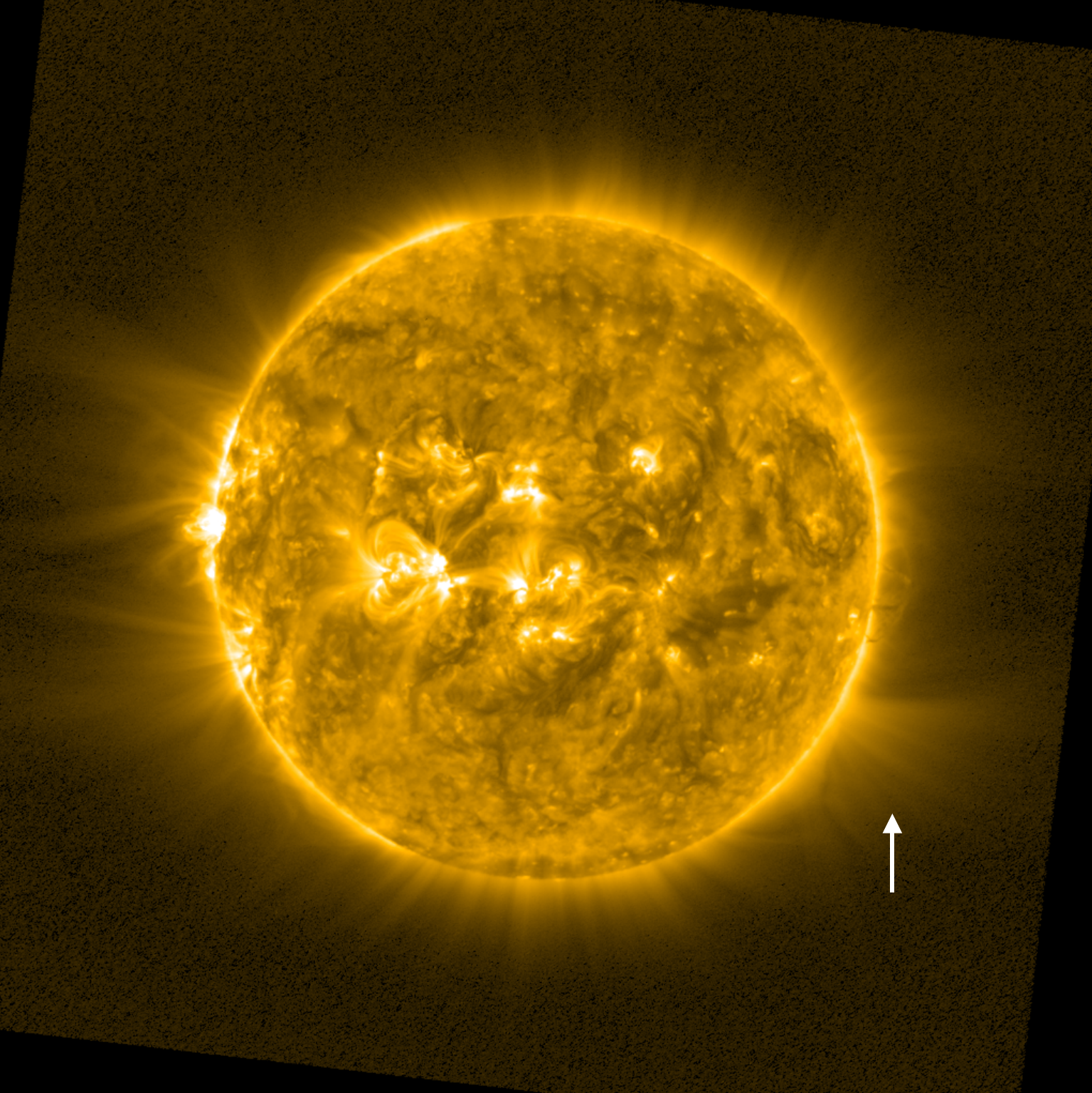}
    \end{interactive}
    \caption{{PROBA2/SWAP 174~\AA\ movie of the flare from the perspective of the Earth. The movie has been rotated to align with solar north. An arrow in the static version of the figure indicates the location of the fan structure visible both from SWAP and EUVI.}}
    \label{fig:SWAP}
\end{figure}

The fan structure observed by STEREO-A is particularly important. Because it extends high into the corona, it was also visible in 174~\AA\ observations from PROBA2/SWAP\footnote{Images from the Solar Dynamics Observatory's Atmospheric Imaging Assembly were unavailable at this time due to a calibration maneuver.}. Like XRT, SWAP detected some eruption-associated evolution, but no specific direct flare signatures. Figure~\ref{fig:SWAP} and the accompanying animation show SWAP's view of the event from onset to decay. For this analysis we prepared SWAP data using the \textsf{SolarSoft~IDL} routine \textsf{swap\_prep.pro}, with its standard calibration settings including its optional PSF deconvolution step to minimize the risk that stray light contaminated the measurements we discuss below.

Both STEREO and SWAP movies reveal an eruption beginning around 12:30~UT, which precedes the onset of the flare brightening itself by a little under two hours. This eruption initially disturbs the fan structure visible both to EUVI and SWAP, but by about 14:30~UT, when the flare irradiance starts to increase, the fan structure appears to be quiescent. The fan's base is anchored at about $120^{\circ}$ heliographic longitude, while the flare is located between about 125--130$^{\circ}$ longitude -- they are separated by a distance of about 50~Mm, and do not appear to interact during the growth of the post-eruptive arcade. The fan is notably closer to the flare itself than other features visible to SWAP and EUVI.

\begin{figure}
    \centering
    \includegraphics[width=0.5\textwidth]{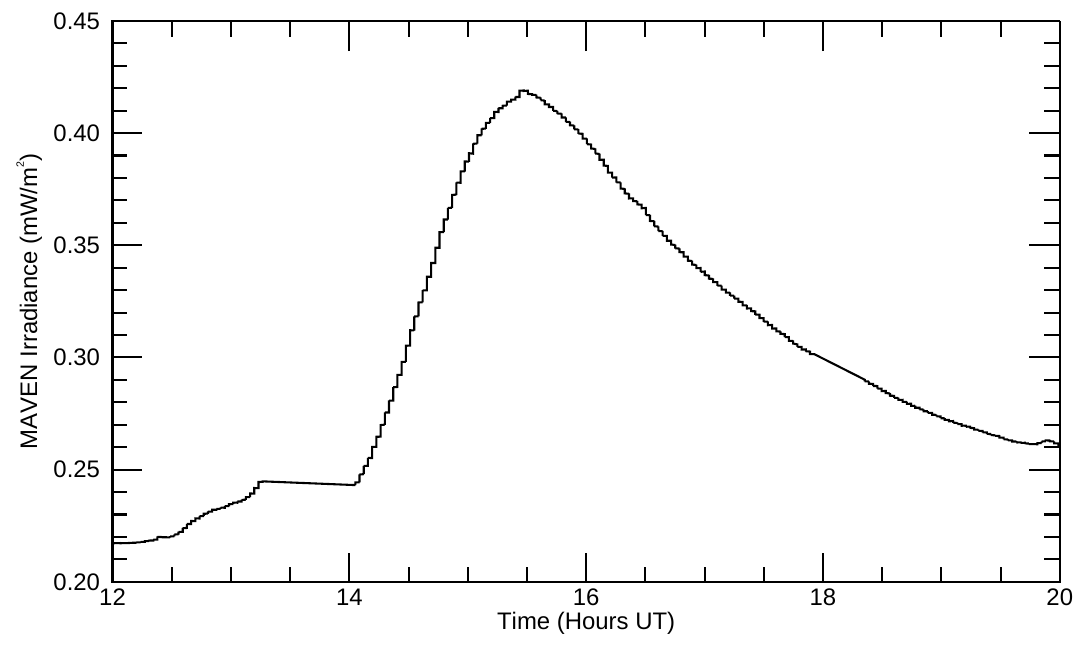}
    \caption{EUV irradiance in MAVEN's EUV Channel~A (about 15--25~nm) at Mars.}
    \label{fig:MAVEN_LC}
\end{figure}

Because STEREO did not observe the evolution of the flare in the 171~\AA\ channel, and was subject to significant saturation in other passbands, we must identify another source of data in order to compare the flare irradiance evolution to features observed by SWAP. MAVEN had a clear view of the flare and records EUV irradiance in the range of about 15--25~nm in its channel~A band, and produced a high-quality light curve for the flare (Figure~\ref{fig:MAVEN_LC}). 

We note that MAVEN's channel~A response contains a second peak in the soft X-ray (SXR) region of 0.1--3~nm. Although this introduces some degree of uncertainty regarding the relationship between MAVEN's observations and SWAP's, in the absence of spectral irradiance measurements it is not feasible to differentiate multiple contributions to the MAVEN observations as a function of wavelength. However, a survey of flare observations made during 2021--2023 suggests that the MAVEN channel~A observations provide a reasonable proxy for the evolution of 174~\AA\ emission in the vicinity of a flare. A full discussion of this survey appears in Appendix~\ref{appendix:survey}.

\begin{figure}
    \centering
    \includegraphics[width=1.0\textwidth]{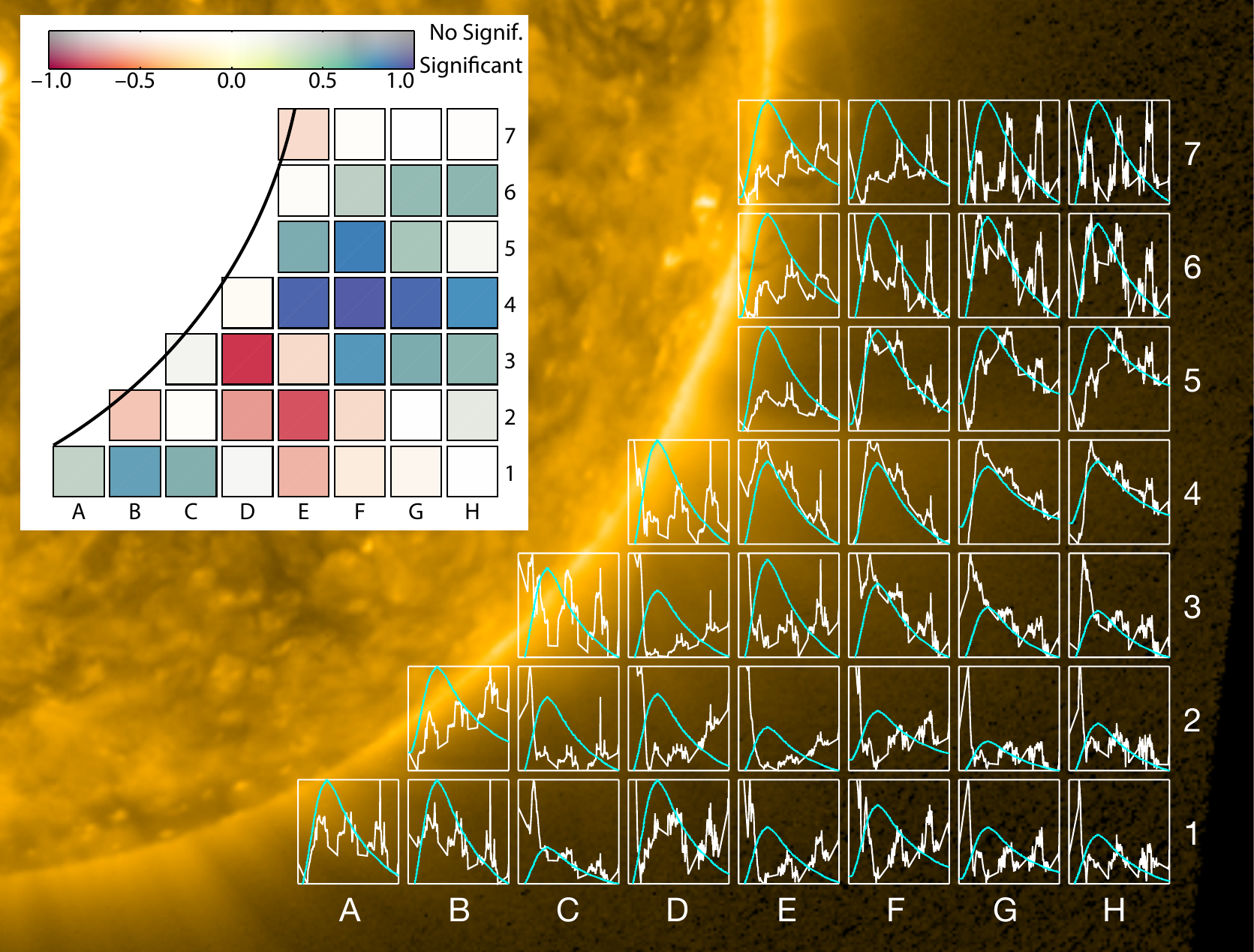}
    \caption{Summary of the comparison of MAVEN EUV irradiance and SWAP brightness as a function of position. Both light curves within each square in the plot are normalized to appear on the same set of axes. MAVEN light curves are cyan, SWAP is white. Rows are labeled with numbers and columns with letters. The inset at upper left shows the Pearson correlation coefficient between the MAVEN and SWAP data for each region, where hue indicates the value of the coefficient, while color saturation indicates the significance.}
    \label{fig:SWAP_Grid}
\end{figure}

A preliminary comparison of the evolution of the flare irradiance and SWAP-reported brightness in several regions of the fan structure showed striking correlation, so we devised a way to compare SWAP and MAVEN observations across the whole region. To do this, we divide the region of the SWAP images near the flare into an $8\times7$~grid, with each square in the grid spanning $40\times40$~pixels (about 125$\arcsec$ on a side). For each square, we compare the SWAP-observed brightness to the MAVEN irradiance (normalizing each light curve, and adjusting the axes to span the min and max of normalized brightnesses during the period of interest). Figure~\ref{fig:SWAP_Grid} summarizes the results of this comparison, revealing the regions where the SWAP brightness was well correlated with MAVEN irradiance, and regions where there was no relationship. (Note that because this event occurred during SWAP's annual eclipse season, several short data gaps appear in the SWAP light curves, corresponding to times when PROBA2's view of the Sun was obscured by the Earth.)

In rows covering structures close to the flare site, particularly rows 3--5 and columns E--H, there is a very strong correlation between the MAVEN light curve and what SWAP observed. In rows covering features that are distant from the flare, or even features where there is no direct line of sight to the flare, there is no correlation, suggesting that RE of EUV illumination originating in the flare behind the limb is causing a brightening of nearby features above the limb.

Comparing the mini-light-curve plots in Figure~\ref{fig:SWAP_Grid} to the fan structure indicated in Figure~\ref{fig:SWAP}, we observe that some parts of the fan (particularly the northerly portion in rows 3--5, columns E--H) show a strong correlation with the flare light curve. The Pearson correlation coefficient for each region provides a quantitative indication of how well correlated the irradiance evolution observed by SWAP and MAVEN really is. The coefficients show strong and significant correlation in the rows noted above, as well as relatively strong negative correlation to the southwest. As we discuss below, this region is not well illuminated by the flare irradiance, so we do not expect a correlation between the SWAP and MAVEN light curves in this region. The negative correlation may be the result of a coronal dimming related to the eruption (note the strong dropoff and slow recovery in the SWAP light curve in square E2, for example). While the evidence is not by itself conclusive, it demonstrates clearly that regions in the corona that are not illuminated by the bright flare do not respond in the same way as those that do. Is there anything in the nature of the fan that could explain this? The answer lies in the geometric structure of the fan itself.

\begin{figure}
    \centering
    \includegraphics[width=0.9\textwidth]{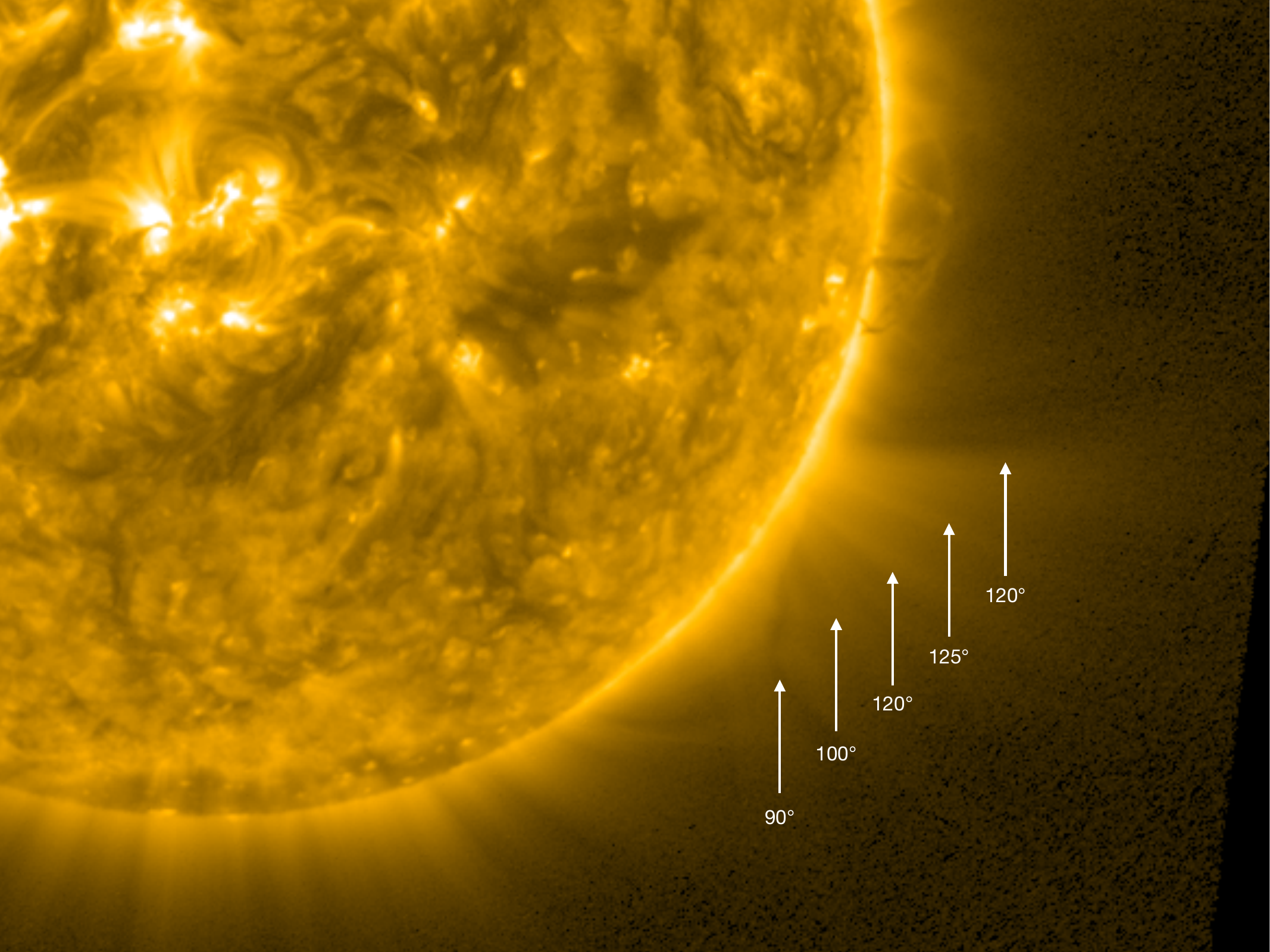}
    \caption{Reconstructed heliographic longitudes of features within the fan structure we studied in this paper. The more southerly part of the fan is considerably west of the northerly portion.}
    \label{fig:Fan_Long}
\end{figure}

Using the routine \textsf{scc\_measure.pro} on EUVI and SWAP observations we estimated the three-dimensional structure of individual features within the fan. We identified several prominent striations in the fan visible to both instruments and used the program to triangulate their position in three-dimensional space \citep[for an example of how this process works, see the discussion in][]{Thompson2009}. We found that, although the fan appears to lie in the plane of the sky from both vantage points, its southerly part, in fact, bends towards the Earth, and away from the flare, as indicated by the arrows in Figure~\ref{fig:Fan_Long}. The southern portion is about $30^{\circ}$ west of the flare, while the northern part is within a few degrees of the flare site itself. This means that, particularly at low heights (columns A--D) the southerly portion lies beyond the horizon from the perspective of the flare, and thus cannot be illuminated by the flare-associated brightening detected by MAVEN. Because of this, no RE due to flare emission can occur in the southerly part of the fan.

\section{A Proof-of-Concept Calculation}\label{sec:test_calculations}

Although this unique observation is strongly suggestive of the presence of resonantly scattered emission in a bright flare, an estimate of this process is necessary to validate its physical plausibility. Here we describe our effort to include sources of RE into a framework for computing synthetic EUV emission, and we illustrate how line-of-sight (LOS) integrated emission can be enhanced by this process.

\subsection{Accounting for the Resonant Excitation in EUV Line Emissivities}

For an in-depth review of the history, theory, and assumptions involved in calculating the emission of optically thin EUV and X-ray emission lines, we refer the reader to the comprehensive review by \citet{delzanna2018a}. For forward modeling such lines, it often is helpful to define the integrated intensity, or radiance of an optically thin coronal EUV emission line at wavelength $\lambda_{ji}$, observed along the line of sight (LOS) as
\begin{equation}
I_{\lambda_{ji}} = \int_{s}G(n_e, T_e,F_\lambda)\ n_e n_H\ ds\quad [\text{erg}\ \text{cm}^{-2}\ \text{s}^{-1}\  \text{sr}^{-1}],
\end{equation}
where $T_e$ is the electron temperature, $n_e$ is the electron number density, $n_H$ is the hydrogen number density, $s$ is the coordinate along the LOS, $j$ and $i$ are the level indices of the atomic transition, and $G(n_e, T_e,F_\lambda)$ is the ``contribution function.'' 

Here we have also added the irradiance of external sources at this point in space, integrated over the spectral line profile, $F_\lambda$ ($\text{erg}\ \text{cm}^{-2}\ \text{s}^{-1}$), as an input to the contribution function to indicate that photoexcitation is also considered. Expressed this way, the local emissivity is split between the contribution of the LOS density distribution and the complexity of the atomic properties (abundances, ionization balance, radiation environment, and the level population of the specific ion and transition), which are all encoded in $G(n_e, T_e,F_\lambda)$. 

As summarized in \citet{delzanna2018a}, a key part of computing $G(n_e, T_e,F_\lambda)$, is solving for a statistical equilibrium for the level population of a given ionic charge-state, and separately calculating the ion abundance assuming ionization equilibrium. For every level this includes accounting for excitation and deexcitation rate contributions from both collisional and radiative processes, which can be solved for simultaneously using modern atomic data and software packages \citep[e.g., the CHIANTI atomic database and analysis package:][]{dere97,delzanna21_chianti10}. For strong coronal EUV lines and typical non-flaring coronal densities ($n_e\!\approx\!10^8$~cm$^{-3}$) it is common practice to assume that collisions generally dominate, and the contribution function is computed as a function of temperature and density.  

We note that the SWAP passband is very broad \citep{Raftery2013} and for typical quiescent coronal temperatures (1--1.5~MK) the emission is dominated by the Fe~\textsc{ix} 171\,\AA, Fe~\textsc{x} 174.5\,\AA, and 177.2\,\AA\ lines.  For our proof of concept, we estimate  the emission due to the Fe~\textsc{x} 174\,\AA\ line, but the other lines would produce very similar contributions.

To add the third-dimensional dependence on the external radiation field, we modify the existing photoexcitation infrastructure in the CHIANTI 10.0.2 atomic database \citep{delzanna21_chianti10} to allow us to manually specify the solid-angle-averaged radiation brightness, $J_\lambda = fF_\lambda/4\pi$, that will be used in the level population solver, where the dimensionless factor, $f$, represents the convolution of the two spectral line profiles (see below). Then we compute $G(n_e, T_e,F_\lambda)$ specifically for the Fe~\textsc{x} 174\,\AA\ line at each point over a $17\!\times\!23\!\times\!32$ grid of $n_e$, $T_e$, and $F_\lambda$. The grid is logarithmically spaced, with coordinates ranging within $n_e=10^{3-10}\ \text{cm}^{-3}$, $T_e=10^{5.3-6.6}$ K, and $F_\lambda=10^{0.76-4.76}$ $\text{erg}\ \text{cm}^{-2}\ \text{s}^{-1}$. $F_\lambda=0$ is also included to provide a table where only collisional excitation is considered.

 It is worth noting that in the case of RE, the exciting radiation comes from the same spectral line, so the photoexcitation rates at a specific point in the corona are determined by the convolution of the spectral radiance of the exciting radiation with the Doppler profile of the absorbing atom. This is commonly done for $\mathrm{Ly\alpha}$ Doppler dimming calculations \citep[e.g., the appendix of][]{dolei19}. 
 
 In our calculations, we simplify things by assuming no flow, thermal and isotropic Doppler velocity of the ion, and a fixed full-width at half-max (FWHM) for the exciting 174\,\AA\ line profile, $\lambda_{FWHM}\!=\!0.036$~\AA, which is intended to represent an average thermal and non-thermal velocity for the quiet Sun. This line profile is convolved with the thermal line profile of the absorbing ion, which varies as a function of temperature, to give the convolution factor, $f(T_e)$. To do this, we adopted a modification of the CHIANTI codes described in Del Zanna (2025). 

\subsection{MHD Model}

With a 3D spectral response table for the Fe~\textsc{x} 174\,\AA\ spectral line computed, we can now use it to examine the role of RE for conditions similar to the flare-illuminated fan structures on 2016~January~6. Because we are specifically interested in characterizing the role of the flare kernel in exciting emission from nearby structures, it is important to have at least a plausible geometry and plasma conditions for a proof-of-concept calculation. For this, we can turn to state-of-the-art 3D magnetohydrodynamic (MHD) models of the global solar corona, in particular those that describe the plasma state with sufficient accuracy to compute EUV and soft X-ray observables that are comparable to observations. 

In Figure~\ref{fig:photex_experiment}a, we show an image of the off-limb radiance of Fe~\textsc{x} 174\,\AA\ computed from one such model with the external radiation field set to zero (collisional excitation only). This particular high-resolution simulation, produced by the Magnetohydrodynamic Algorithm outside a Sphere (MAS) code \citep[e.g.,][]{mikic99,lionello09,mikic18}, was designed to capture the structure and appearance of the solar corona during the 2021~December~4 total solar eclipse\footnote{\href{https://www.predsci.com/eclipse2021}{https://www.predsci.com/eclipse2021}}. It utilizes a Wave-Turbulence-Driven (WTD) approach for prescribing coronal heating \citep{downs16,mikic18} with up-to-date parameters based on favorable comparisons to observations made in the low and middle corona \citep{boe21,boe22}. For our purposes, this simulation was convenient because there were several active regions present in the model, some of which showed visible fan-like structures extending off-limb in forward-modeled Fe~\textsc{x} 174\,\AA\ emission.

To examine this further, we select a single line of sight at 1.2~$R_\sun$ passing through the fan structure, indicated by the red cross symbol in Figure~\ref{fig:photex_experiment}a. In Figure~\ref{fig:photex_experiment}b, we plot the temperature and density extracted along this line of sight as a function of the LOS coordinate, $s$, with the negative axis oriented towards the observer. As we can see, the density grows larger near the plane of sky ($s\!=\!0$), but its distribution is not uniform, with essentially one peak at $s\!\approx\!-0.3~R_\sun$ and another at $s\!\approx\!0~R_\sun$. Similarly the temperature distribution is also non-uniform, showing that the plasma along this particular line of sight is inherently multi-thermal. The overlap of the two curves where the plasma is cool ($\sim$1~MK) and dense ($\lesssim$10$^8$~cm$^{-3}$) near the plane-of-sky effectively localizes the fan emission in space, as indicated with the pink shaded region.

\subsection{Estimating the Sources of Illumination}
\label{subsec:sources}

Using the knowledge of temperatures, densities, and position along the LOS we can now compute emissivities for the various emission mechanisms and sources of illumination. $F_\lambda$ can be determined by knowing the radiance of all external sources as a function of solid angle as seen by a local point in space. If we approximate a source of external radiation as a uniformly bright sphere of radius, $r_0$, then the integral over solid angle for a point at a distance, $d$, is computed as:
\begin{equation}
F_\lambda = I_\lambda \int_0^{2\pi}\int_0^{\sin^{-1}(r_0/d)} \sin{\theta}d\theta d\phi =  2\pi I_\lambda \left ( 1 - \sqrt{1 - r_0^2/d^2} \right ),
\label{eq:rad_to_irad}
\end{equation}
where the limits of integration describe the cone of solid angle centered on the source and ending at the apparent edge of the sphere.

With this spherical approximation, we can divide the sources of external radiation for RE into two pieces, the average contribution of Fe~\textsc{x} 174\,\AA\ emission emanating from the quiet low corona ($r_0\!=\!r_\sun$), and the additional illumination from the bright flare ($r_0\!=\!r_{FL}$). For the quiet corona, we estimate the disk-averaged surface brightness using SDO EUV Variability Experiment (EVE; \citealt{Woods2010}) spectral irradiance measurements \citep{woods12}. Following the choice of \citet{delzanna19} who used 2010~May~16 as a relatively quiet day for measuring spectral irradiances, we integrate the merged EVE Level-3 (version~7, revision~2) 0.2\,nm spectral irradiance data on this day from 173.6 to 175.6\,\AA\ to cover the Fe~\textsc{x} 174\,\AA\ line profile and estimate the total irradiance of the line, $F_{QS}\!=\!0.04655$ erg cm$^{-2}$ s$^{-1}$. Then we obtain the average disk brightness, $I_{QS}=685$ erg cm$^{-2}$ s$^{-1}$ sr$^{-1}$, by inverting Eq.~\ref{eq:rad_to_irad} and using the distance from the Sun to Earth. 

To compute the radiation contribution from the flare kernel, we require a heuristic estimate of its size and brightness. First, via inspection of the STEREO-A/EUVI 171~\AA\ image sequence, we estimate a rough angular size of the flaring region of $3.8\times10^{-8}$\,sr by summing the areas of significantly brightened pixels in the images. Then we compute the size of the sphere that would subtend this solid angle as seen by STEREO-A, which gives an approximate flare kernel radius of $r_{FL}\!\approx$0.023~$R_\sun$.

Given that the MAVEN channel A irradiance passband covers a number of coronal lines that should ostensibly all evolve during the flare, it is not possible to isolate the increase solely in Fe~\textsc{x} 174\,\AA\ irradiance during the flare. Instead, based on the analysis in Appendix~\ref{appendix:survey}, we can make the simple assumption that the increased Fe~\textsc{x} 174\,\AA\ irradiance roughly follows the temporal evolution of the flare in this passband and is some small fraction of the average full-disk quiet-sun irradiance that we obtained above. 

Previous studies suggest that the irradiance of 1~MK lines varies at most by a few percent during the peak of the X-ray emission of large flares \citep[see, e.g.,][]{DelZanna_Woods_2013} although the localized increase in the vicinity of hot flare loops can be much more significant (see the discussion in Appendix~\ref{appendix:calculations}). Free-free continuum contributes significantly to the increase in emission in broad-band  EUV imagers for large flares, but its relative contribution compared to line emission in the 171/174~\AA\ range have not been well characterized empirically due to the confounding effects of coronal dimming on irradiance measurements \citep[e.g.,][]{Mason2014}. \citet{Milligan2013} reported continuum contributions of 8--43\% in AIA channels with hot EUV lines. \citet{ODwyer2010} estimated continuum would contribute about 23\% the total flare-related radiance in the AIA 171~\AA\ channel. In any case, the convolution of the disk emission and absorption profile is such that the contribution from the continuum would be very small, compared to that of the line.

To better constrain the specific relationship between MAVEN's observations and the evolution of EUV line emission, we carried out a survey of flare observations using the \textit{GOES} Solar Ultraviolet Imager \citep[SUVI;][]{Darnel2022}. Our results showed that for typical large flares the increase in the irradiance at around 171~\AA\ is between a few percent and about 10\% (see Appendix~\ref{appendix:survey}). For illustrative purposes within our simple model, we assume that the flare adds a 10\% contribution to the total irradiance, which is on the larger end for spectral irradiance variations in Fe~\textsc{x} lines both according to our survey and as measured by SDO/EVE during flares \citep{hock12_thesis, mason19_catalog}, but from the XRT observations of flare-associated SADs and the very fast CME it produced, the event appears to be a very energetic one and the flare likely to be very bright. Assuming continuum contributions to SWAP's 174~\AA\ channel are consistent with other EUV observations of flares, we assume 50\% of the increase we observe results from continuum, and thus the available irradiance for our model is 5\% of the pre-flare value. (50\% continuum contribution is likely an overestimate, but provides a strong upper limit on the contribution.)

Assuming an increase in total emission line irradiance of 5\%, the flare brightness integrated over the line profile is simply the fractional contribution multiplied by $I_{QS}$ and the ratio of the solid angle of the solar disk to that of the flare as seen by STEREO-A. This gives $I_{FL}=6.65\times\text{10}^4$ erg cm$^{-2}$ s$^{-1}$ sr$^{-1}$, or $\sim\!97$ times larger than the average quiet-sun brightness.

\subsection{Modeling the Resonant Excitation Contributions}

\begin{figure}
    \centering
    \includegraphics[width=0.99\textwidth]{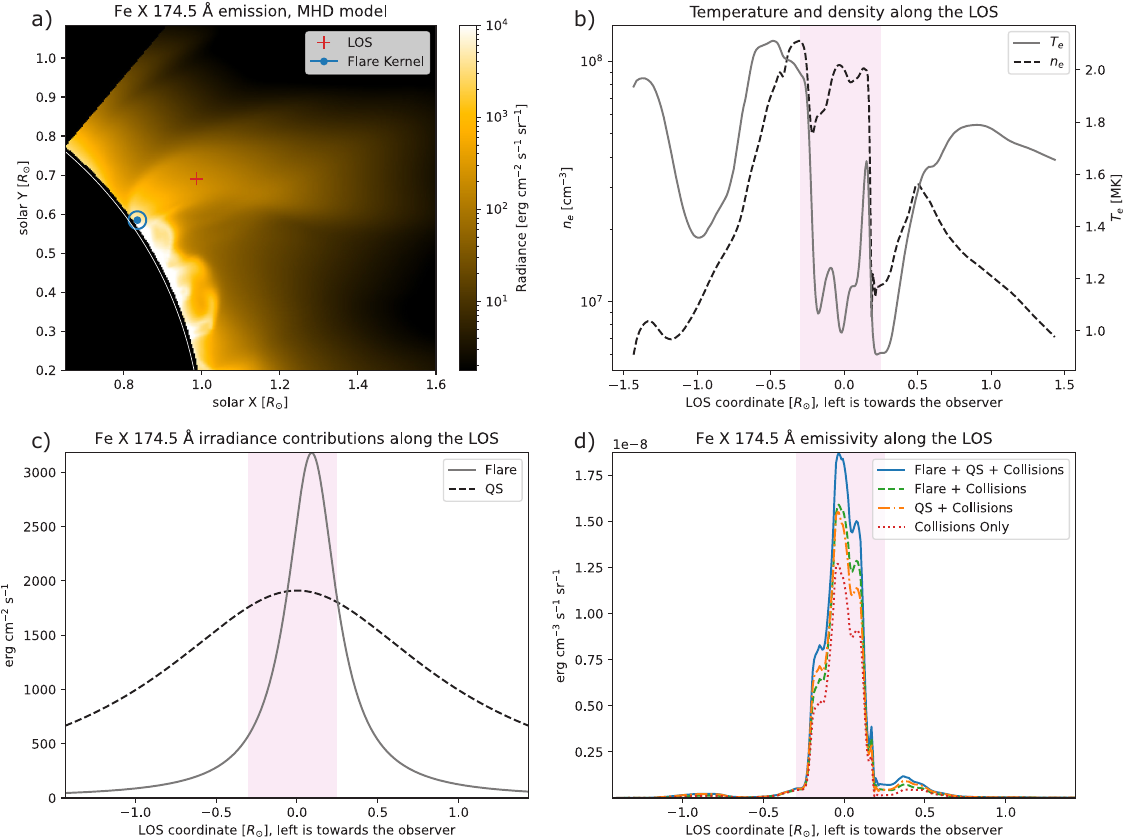}
    \caption{Results for the test calculations of RE with an MHD model. a) Context image showing forward-modeled Fe~\textsc{x} 174.5~\AA\ emission generated from the model assuming no photoexcitation. The red crosshair indicates the selected LOS, which intersects a fan structure at 1.2~$R_\sun$. The blue circle indicates the location and assumed size of the flare kernel, which was placed 5$^\circ$ behind the plane of sky. The remaining three panels show properties along this LOS in plane-of-sky coordinates. b) Electron density and temperature; c) Irradiance contributions from the quiet-sun (QS) and the peak of the flare; d) Local emissivity computed for all four different photoexcitation assumptions. The pink shaded region indicates the portion along the LOS where most of the emission is formed.}
    \label{fig:photex_experiment}
\end{figure}

With our response tables, MHD model, and brightness estimates in hand, we can finally calculate the different contributions to Fe~\textsc{x} 174\,\AA\ emission. The first step is to compute the irradiance of each brightness source at points along the LOS. For the quiet-sun source, the distance needed for Equation~\ref{eq:rad_to_irad} is simply the radial position of the point, but for the flare contribution we must compute the distance relative to the flare kernel in 3D space. For this experiment  we place the flare $r_{FL}$ above the surface, at the same latitude as the intersection of the LOS with the plane of sky, and 5$^{\circ}$ longitude behind the plane of sky. This position is indicated in Figure~\ref{fig:photex_experiment}a.

Figure~\ref{fig:photex_experiment}c shows the irradiance contribution of each source along the line of sight. As expected, the quiet-sun illumination peaks at the center of the plane of sky but falls off relatively slowly. The flare illumination is far stronger near the flare ($\sim$1.7$\times$), but falls much more quickly as the apparent size of the flare as seen by points along the LOS diminishes rapidly.

Next we pass the temperature, density, and illumination at every point along the LOS to the 3D contribution function table get the local emissivities ($G(n_e, T_e,F_\lambda) \frac{n_h}{n_e} n_e^2 $). By varying which illumination sources are considered, we can create different permutations of the emissivity, which are shown in Figure~\ref{fig:photex_experiment}d. The first considers only collisional processes (red), then we add photoexcitation from either the quiet-sun (yellow), the flare (green), or both (blue). All emissivity curves look similar, showing a strong degree of localization where the temperature is low and densities high within the fan structure, however those including illumination are also separated from one another, indicating a non-negligible contribution from both sources of photoexcitation. 

Next we integrate the emissivity curves along the LOS to obtain what would be observed, giving radiances of 223, 290, 291, and 355~erg\,cm$^{-2}$\,s$^{-1}$\,sr$^{-1}$, respectively, for each permutation. Comparing the first two numbers we see that the quiet-sun contribution of RE to the fan brightness is roughly 30\% of the collisional component. This is interesting because it indicates that, even in the absence of a flare or nearby bright region (i.e., quiescent conditions),  RE may contribute more than traditionally expected to the emission of bright structures observed off-limb (at least for strong, bright lines like Fe~\textsc{x} 174\,\AA).

Finally we see that the illumination by the flare indeed produces a significant increase in the brightness of the fan structure, with the full calculation being about 22.5\% brighter than when only including quiet-sun illumination. This shows quite clearly that a significant enhancement of emission within a fan structure is possible in the localized region surrounding the flare. Of course, as illustrated by Figure~\ref{fig:photex_experiment}c, this factor will depend strongly on the distance of the fan structure to the flare kernel.

Despite the heuristic flare estimates and the simplifying geometric assumptions, this experiment shows that the contribution of RE can be non-negligible -- or even significant -- off-limb given sufficient proximity to bright structures. By the same token, this analysis illustrates the inherent complexity of the RE problem, where the specific properties of narrow multi-thermal structures along the line of sight must be convolved with their view of the surrounding radiation field at a given wavelength. In that sense, this calculation should be considered purely a proof of concept that attempts to show how the off-limb enhancement observed during this event could be at least plausibly consistent with RE. If we had direct measurements of the disk emission and of the line-of-sight distribution of densities and temperatures, more accurate estimates would be possible. Our estimates 
would change depending on assumptions; a more exhaustive exploration of this concept for all lines of sight and different models or views is left to future work.

\subsection{Alternative Explanations}\label{sec:alternatives}

The spatial and radiometric correspondence in our data is suggestive, and our simple model demonstrates the possibility that RE of illumination from a flare could significantly influence the overall radiance of off-limb structures in the low and middle corona. Nonetheless, we must exclude other possibilities that could explain the correspondence in brightness evolution in the flare and fan structure. Two obvious explanations are that the relationship is purely coincidental -- a superficial similarity with no causal link -- and that the same process that heats the flare and causes its evolution acts on the fan as well.

Although separated features within the corona could, in principle, exhibit the same evolution purely by coincidence, it is unlikely that such striking correspondence -- in both onset, rate of rise and fall, and duration, as we observe in row 4, columns E--H in Figure~\ref{fig:SWAP_Grid} -- would occur entirely by chance. Moreover, we would not expect a systematic coupling across such a large-scale structure as the one we study here to have a random origin. It is more likely the systems are connected somehow, so we examine these other possible explanations.

A more likely possibility is that both the structure and the flare itself are heated by the same magnetic reconnection process, and thus experience the same evolution. In flare models, reconnection generates and heats post-eruptive loops by extracting energy from the surrounding magnetic field. The footpoints of recently reconnected loops mark the locations where the process of reconnection is connected to structures within the corona. Models of flares including thermal conduction \citep[e.g.,][]{Yokoyama2001, Seaton2009} suggest that a flare can also heat a region of plasma surrounding the flare site, generating a ``thermal halo,'' but that this region is confined to a narrow band in the immediate vicinity of the flare current sheet. This heating is generally isolated to the region where magnetic field lines connect across the slow shocks (i.e., Petschek shocks) that bound the high-speed reconnection outflow jets.

In the present case, the base of the fan structure is clearly separated from the flare site. Based on computed longitudes, the minimum separation is $>$50\,Mm, and even at their widest extent, the flare ribbons are well separated from the fan. This separation strongly suggests that heating due to the flare does not directly drive the behavior of plasma in the fan. More importantly, even if it did, it is not likely that flare heating would actually result in an increase in the brightness of this structure as we observe here. Instead, heating is likely to shift the emission out of SWAP's 174\,\AA\ passband, which has a peak temperature response around 0.85~MK and is only weakly sensitive to higher temperatures \citep{Raftery2013}. 

Cooling processes within the fan are unlikely to operate with the same efficiency as in the dense, short loops associated with the flare, resulting in different temporal evolution in the light curves from the fan. Even more importantly, because thermal conduction within single magnetic structures in the corona is extremely efficient, we can expect that any changes in temperature along a single field line of the fan would be observed to be correlated across the whole structure, without respect to altitude. This stands in contrast to the behavior in Figure~\ref{fig:SWAP_Grid}, rows~3 and 4, where low-altitude parts of the fan, for which the flare kernel is largely over the horizon, do not behave the same as the high-altitude parts, for which the flare is visible.

There are other more complex possibilities -- the initial eruption does disturb some structures in the vicinity of the fan, for example -- so it remains possible some connection exists that could explain this relationship. But there are arguments against this possibility as well.

First, the animations of the flare (Figures~\ref{fig:EUVI-195-movie} and \ref{fig:SWAP}) show clearly that the motions and changes of the fan's physical structure have stopped by 15:00~UT, very early in the evolution of the flare light curve itself. Any brightness evolution that occurs because of the dynamics of the initial eruption process are likely to be completed at this time.

Furthermore, if the eruption somehow \textit{is} connected to the overall correspondence, why does it affect the fan features in the north, but not the south? Likewise, why would it affect fan-associated features at larger heights, but not in the low corona? Given the close relationship between the fan's geometry and the locations where this correlation between flare and fan brightness occurs (or does not occur), RE of the EUV illumination generated by the flare offers a much simpler and more self-consistent explanation.
 
\subsection{Implications}\label{sec:implications}

\begin{figure}
    \centering
    \includegraphics[width=0.5\textwidth]{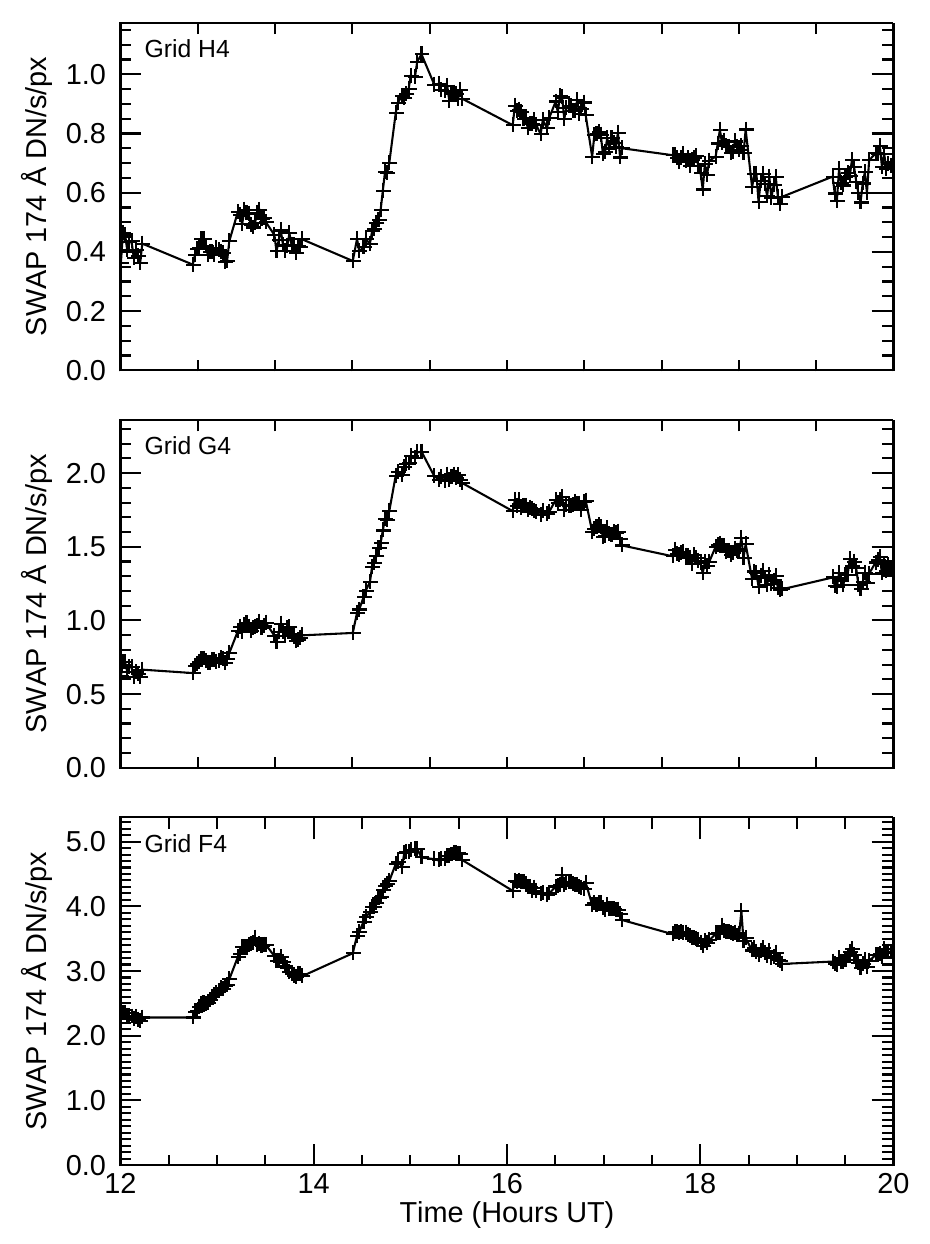}
    \caption{Time-normalized SWAP counts in selected grid squares from Figure~\ref{fig:SWAP_Grid}.}
    \label{fig:swap_raw}
\end{figure}

Figure~\ref{fig:swap_raw} shows the average time-normalized counts per pixel in three of the grid squares plotted in Figure~\ref{fig:SWAP_Grid}. Resonant excitation of flare emission apparently contributes significantly to the emission in these squares between about 14:30--19:00~UT, but naturally it does not operate before the start of the flare. Thus we can estimate the fraction of emission arising purely from RE by comparing the flare part of the light curve to the background levels before the flare begins. In squares H4 and G4, more than 60\% of the emission apparently comes from RE, while a little lower in the corona, in square F4, about 40\% of the emission comes from RE. 

In fact, this probably represents a lower limit on the total contribution due to RE, since this estimate depends on our assumption that all of the pre-flare emission was collisionally excited, which we know from our simple model in Section~\ref{sec:test_calculations} is probably not accurate. Nonetheless, the increases we observe in the light curves are more or less consistent with the model's predicted increase due to scattering of flare emission specifically. We conclude that it is very likely that RE played an important role in generating the emission we observe from the coronal fan during this eruption.

It remains almost universally accepted that the only significant emission from the corona in EUV results from collisional excitation. This assumption has important implications for a variety of measurements techniques used to characterize coronal plasma -- perhaps most notably differential emission measure (DEM) analysis \citep[see, e.g.,][]{Cheung2015, Plowman2020}. DEM techniques allow observers to determine the amount of emitting material -- or emission measure -- of different temperatures that lies along the line of sight of any pixel in a set of EUV observations, and is widely used to determine temperatures in phenomena ranging from coronal loops to solar flares. But what if the key assumption of these techniques -- that the emission mechanism in the observations is well known -- is not correct? 

Although, like collisionally excited emission, resonantly scattered emission is sensitive to the temperature distribution in the plasma where it originates, it scales linearly with electron density, while collisional emission scales as the density squared. Adaptations to the DEM analysis techniques like those implemented for visible and infrared lines by \citet{Boe2023} must be implemented for EUV lines as well, particularly for middle-corona features. An important benefit of new diagnostic techniques that harness the two-component dependence of emission on density is that they may provide much stronger constraints on the plasma density itself, especially in structures that can be clearly localized along a line of sight.

As both our observations and modeling show, during flares, DEM calculations sensitive to RE in EUV might prove to be more challenging than during quiescent conditions. As we see in Figures~\ref{fig:Fan_Long} and \ref{fig:photex_experiment}, the contribution from RE will depend on the three-dimensional geometry of the region. While DEM tools can probably be adapted to account for this in specific cases, there are instances -- such as the one presented here -- where it will be impossible to accurately account for the incident illumination due to a flare, and others where it will not be possible to determine the 3D geometry of a region. In these cases, it will be critical to remain aware of the possible limitations of DEM analysis tools, particularly at larger heights in the corona.

\section{Summary and Conclusions}\label{sec:conclusions}

Attempts to characterize the contribution of RE to the global coronal EUV brightness during solar flares are often hampered by the challenge of disentangling instrumental scattered light from processes in the solar atmosphere. Here we have presented SWAP observations that avoid that problem. Even though the flare is completely occulted from SWAP's point of view, there is a close correspondence between the brightness evolution observed in a coronal fan near the flare site and EUV irradiance measurements of the flare from MAVEN.

We conclude that RE is the most likely explanation for this correspondence, and that RE in a bright flare such as this one could contribute up to 60\% of the total observed brightness in the fan, particularly at larger heights.

These observations have important consequences for studies of emission in the corona, both locally during flares, and globally during quiescent conditions, particularly for structures at large heights. In particular, there are consequences for structures in the middle corona, above about 1.5~R$_\sun$, where emission due to collisional excitation falls precipitously. However, new developments in DEM analysis techniques that account for collisional and resonantly excited emission can overcome these challenges once they are adapted for EUV observations.

There are also important implications for studies of the middle corona in EUV more generally. Our observations show that the flare approximately doubles the EUV irradiance of the Sun, and causes a corresponding twofold increase in emission in the vicinity of the flare. Given this increase, it is likely that even before the flare a nontrivial fraction of the emission in the fan also originated from RE. At larger heights, where the collisional excitation does not operate efficiently, even a small contribution by RE might represent a significant fraction of total emission. Such a conclusion would be consistent with observations by \citet{Seaton2021} and modeling by \citet{Gilly2020}, so predictions of the brightness of the middle corona that rely exclusively on collisional excitation could be significant underestimates.

In fact, this is probably good news for instruments that observe the middle corona in EUV, including the EUV Full-Sun Imager \citep[FSI;][]{Rochus2020, Auchere2023} on \textit{Solar Orbiter}, the Sun Coronal Ejection Tracker \citep[SunCET;][]{Mason2021} CubeSat, currently under development, and the \textit{EUV CME \& Coronal Connectivity Observatory} (ECCCO) mission concept currently in Phase A. If RE contributes significantly to emission in the middle corona, all of these instruments will outperform pre-flight expectations derived using the assumption that the only source of emission in the middle corona is collisional excitation. Indeed, observations of a prominence eruption in FSI's 304~\AA\ passband, which remained bright beyond 6~R$_{\sun}$, by \citet{Mierla2022} have been attributed to RE.

Multi-perspective observations using observatories located near Earth, especially those including spectroscopic capabilities such as ECCCO and the planned Extreme Ultraviolet High-Throughput Spectroscopic Telescope (EUVST) and Multi-slit Solar Explorer (MUSE), complemented by observations from elsewhere, such as from Solar Orbiter or the planned Vigil mission to the L5 Lagrange point, should offer opportunities to better constrain the role of various emission mechanisms, by providing additional constraints on the contributions to coronal emission in future events with similar geometry to this one. The resulting improvements in models of the middle corona that more accurately capture all sources of emission will yield significant improvements in performance metrics and requirements for future instruments. Likewise, new spectroscopic observations, such as those from ECCCO, will finally allow us to fully characterize emission from bright structures in the corona and their relationship to the radiometric evolution of nearby features. 

\begin{acknowledgments}
This work was motivated by discussions during the formulation of the EUV CME and Coronal Connectivity Observatory (ECCCO) mission concept, and we thank the entire ECCCO team for productive discussions that identified this interesting event and motivated the paper. We thank Chris Lowder for helpful conversations during the development of various figures. DBS, MJW, CD, and AC acknowledge support from NASA’s Heliophysics Guest Investigator program, grant 80NSSC22K0523. 
GDZ acknowledges support from STFC (UK) via the consolidated grant to the atomic astrophysics group  at DAMTP, University of Cambridge (ST/T000481/1). 
We acknowledge useful discussions at the International Space Science Institute (ISSI) in Bern, through the ISSI International Team project \# 23-572 on {\it Models and Observations of the Middle Corona}.
SWAP is a project of the Centre Spatial de Liege and the Royal Observatory of Belgium funded by the Belgian Federal Science Policy Office (BELSPO). Hinode is a Japanese mission developed and launched by ISAS/JAXA, with NAOJ as domestic partner and NASA and STFC (UK) as international partners. It is operated by these agencies in co-operation with ESA and NSC (Norway).
\end{acknowledgments}

\facility{PROBA2 (SWAP), GOES (XRS), Hinode (XRT), STEREO (EUVI)}

\appendix 

\section{Survey of EUV Flares and Relationship to MAVEN}
\label{appendix:survey}

The MAVEN EUV monitor's channel A, used in this study, has two separate bands of sensitivity: 0.1--0.3\,nm and 16--22\,nm. Disentangling contributions in these two regions is not straightforward. If a significant contribution to the integrated total solar irradiance due to a flare comes from soft X-rays, then we might not expect the MAVEN EUV channel A measurements to be a good indicator of EUV emission in the vicinity of the flare, and thus we cannot estimate how much irradiance would be available to drive scattering during the event of 2016 January 6 from MAVEN data alone.

Likewise, from the data available, we also cannot assess how much the EUV emission alone increased during the event. The flare is not observed from the Earth's perspective, and STEREO/EUVI observations are significantly saturated in the neighborhood of the flare, so a full accounting of the increase in the EUV flare emission is not possible. \citep[A study with SWAP showed that the vast majority of flare-related radiance is not accounted for even during moderate events; see][Figures~6.7 and 6.8.]{Bonte2013} Therefore, we turned to other data -- and other flares -- to assess how well correlated we might expect 174~\AA\ emission (Fe~\textsc{ix} and Fe~\textsc{x}) and MAVEN light curves to be, and how much of an increase in emission we would expect on both global and local scales.

We use Level-2 data from the \textit{GOES} Solar Ultraviolet Imager \citep[SUVI;][]{Darnel2022}, which are high-dynamic-range images generated from multiple exposures and filter combinations, in which there is essentially no saturation, to compare to MAVEN EUV measurements and assess the relationship between local and global irradiance in the relevant spectral range to higher temperature emission that might be better correlated with MAVEN's channel A data.

We examined 22 different solar flares observed by SUVI between 2021 and 2023, including two for which there are also MAVEN measurements: an X1.0 flare on 2022~October~2 and X1.1 flare on 2023~February~11. {We note that, because the data available suggest the event we observed here was a large and energetic flare, for our survey we consider flares of class X1.0 and above. Although the irradiances involved would intrinsically be less for a weaker event, flare irradiance varies across all classes , and the overarching relationships we identify here should hold for an M-class flare or potentially even weaker event.}

\begin{figure}
    \centering
    \includegraphics[width=0.95\textwidth]{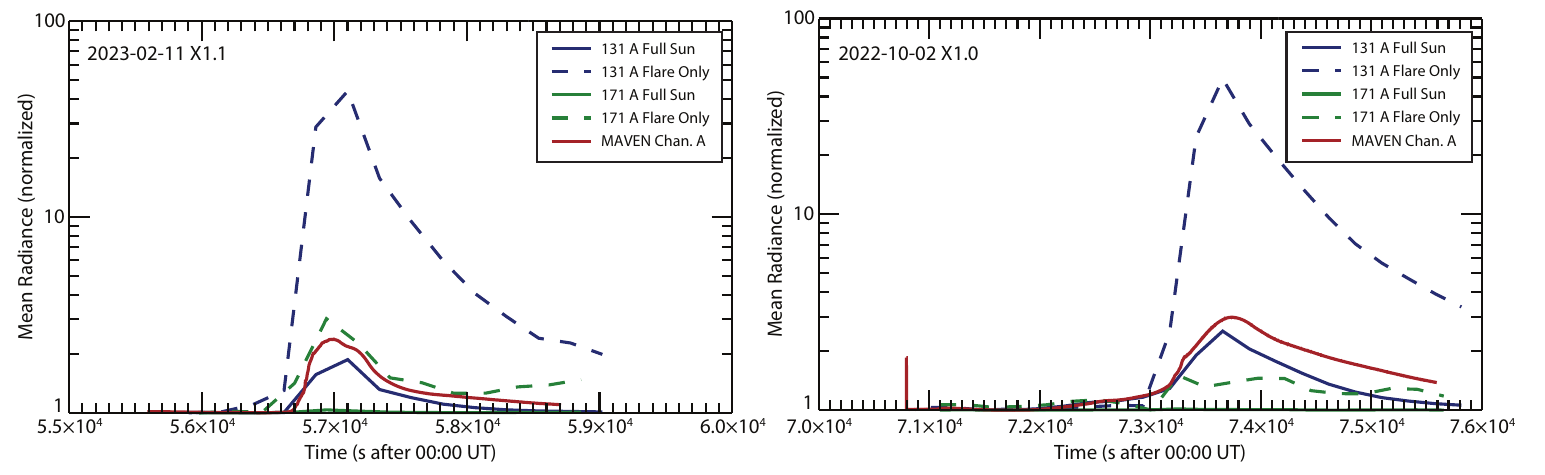}
    \caption{Normalized SUVI 171 and 131~\AA\ light curves for the full Sun and the flaring region, compared to MAVEN channel A irradiance for two flares simultaneously observed by the two instruments. MAVEN data are corrected for light travel time to Mars vs. Earth. Data are normalized such that the plot reflects the overall increase in brightness from the preflare conditions.}
    \label{fig:SUVI_MAVEN}
\end{figure}

\begin{figure}
    \centering
    \includegraphics[width=0.95\textwidth]{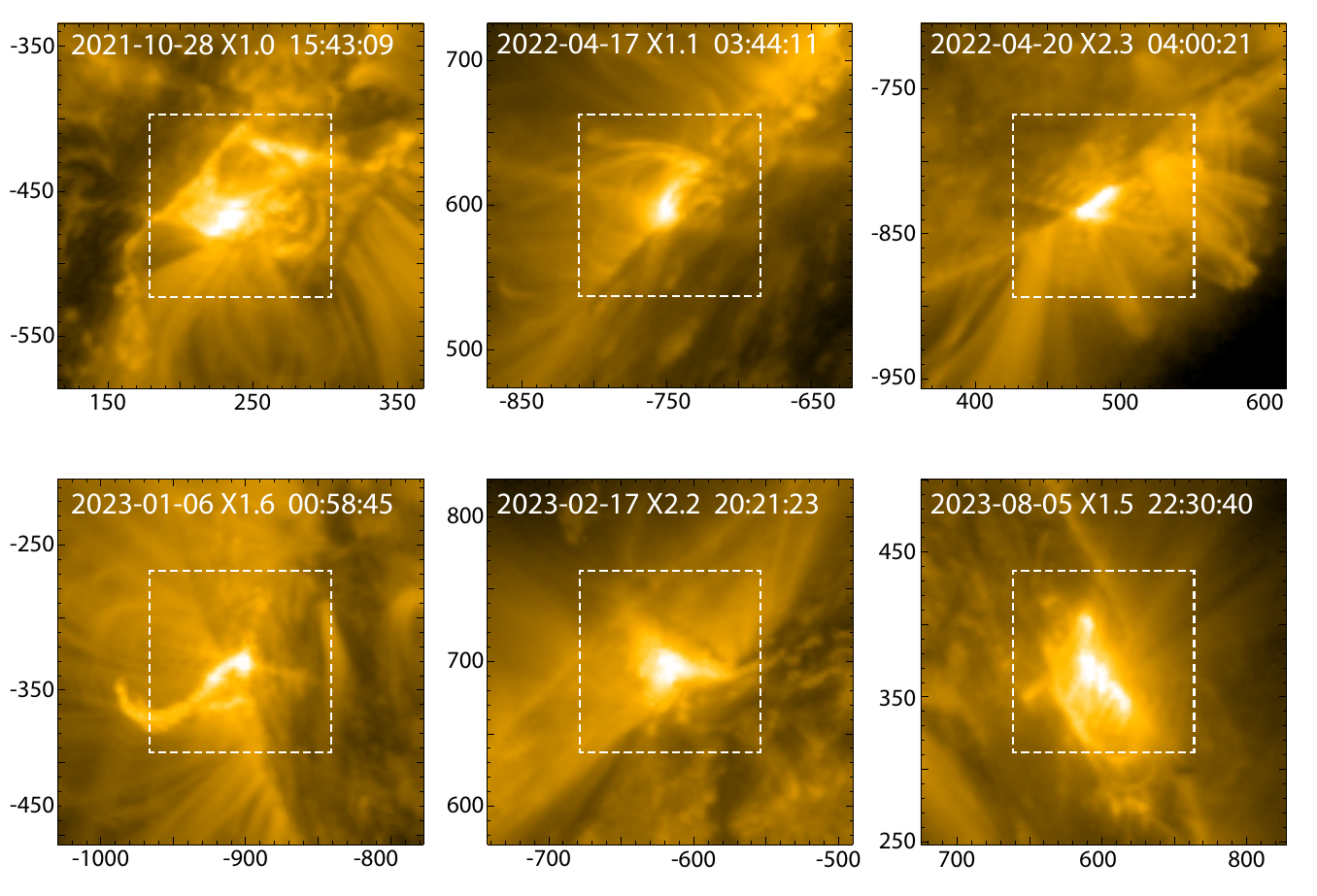}
    \caption{Log-scaled SUVI 171~\AA\ images of flare peaks for selected events included in this survey. The dashed white box shows the region used to estimate the local increase in irradiance due to the flare as discussed in this Appendix. Axis labels are helioprojective coordinates in x- and y-directions, measured in arcseconds from Sun center.}
    \label{fig:flare_morphology}
\end{figure}

Figure~\ref{fig:SUVI_MAVEN} shows a comparison between SUVI 171 and 131~\AA\ radiance, both integrated over the full Sun and in a $125'' \times 125''$ region surrounding the brightest part of the flare, and MAVEN channel A total solar irradiance. Data are normalized to show how much the brightness increases from the preflare conditions. The size of this region far exceeds the size of the flare itself (see Figure~\ref{fig:flare_morphology}), so the local increase in radiance from the flare is a significant underestimate of the fractional contribution from the flare radiance itself. However, selecting a larger region ensures no flare features are inadvertently excluded from our analysis.

In the case of the 2023 flare, the brightness increase in the 171~\AA\ channel in the neighborhood of the flare is well correlated with both the MAVEN measurements and with the SUVI 131~\AA\ light curves. The correlation of 171~\AA\ brightness to both 131~\AA\ and MAVEN is poorer in the case of the 2022 event, though the timing of the brightness increase matches across all channels. However, 131~\AA\ full-Sun irradiance increases are well correlated to MAVEN in both cases, suggesting that 131~\AA\ emission in SUVI is a reasonable proxy to MAVEN channel A irradiance.

\begin{figure}
    \centering
    \includegraphics[width=0.95\textwidth]{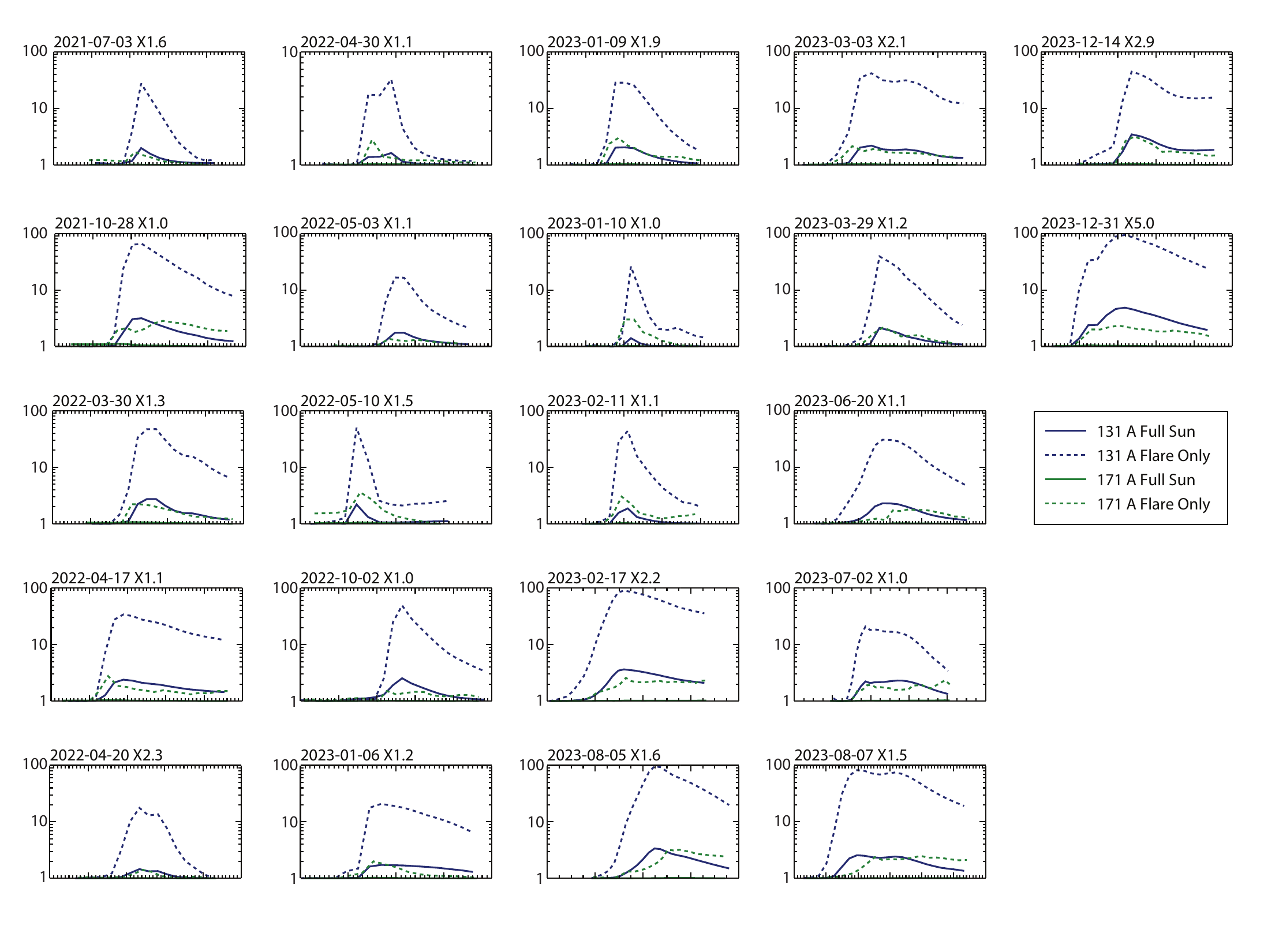}
    \caption{Normalized SUVI 171 and 131~\AA\ light curves for the full Sun and the flaring region for large flares during the period of 2021--2023. The 171~\AA\ increase in full-Sun radiance of a few percent is too small to be seen in most of the plots above.}
    \label{fig:plot_grid}
\end{figure}

Figure~\ref{fig:plot_grid} shows the comparison of all 22 flares we examined during our study period. As in the two flares above, the localized increase in brightness of emission in the 131~\AA\ channel is greater than a full order of magnitude. The local increase in 171~\AA\ emission is roughly equal to the global increase in 131~\AA, consistent with the behavior shown in Figure~\ref{fig:SUVI_MAVEN}.

An important parameter for our model in Section~\ref{subsec:sources} is the global increase in irradiance in the 171~\AA\ band due a flare. For our sample of flares, the integrated irradiance increases by an average of $3.5\pm2.4$\% (median: 2.5\%), with a minimum increase of 1.1\% and a maximum of 11.7\%. Global irradiance increases include the balance of increased brightness in flare-related structures and decreased brightness due to large-scale dimming \citep[see, e.g.,][]{Mason2014} and therefore may represent an underestimate of the localized increase in brightness due to the flare itself. For our simple model we assume a net increase in irradiance of 5\% as the result of a flare, which is above the mean of the population in this survey, but is not inconsistent with our findings here.

In the absence of saturation-free spectral irradiance measurements of the Fe~\textsc{ix} and Fe~\textsc{x} lines that dominate SUVI's 171~\AA\ and SWAP's 174~\AA\ channels, only proxy measurements, such as MAVEN's EUV channel A, are available to track the irradiance of the flare presented in this paper. However, our study of 22 flares suggests that for large flares the evolution of global irradiance of hot lines observed in SUVI's 131~\AA\ channel is a reasonably proxy for the evolution of irradiance for 171~\AA\ emission near the flare. Further, the global irradiance evolution in the 131~\AA\ is well correlated with the MAVEN EUV channel A irradiance for the two flares where overlapping observations are available. We therefore conclude that MAVEN EUV channel A measurements are good indicators of the localized increase in brightness in SWAP's 174~\AA\ channel as the result of a flare.  We note that this relationship is a \textit{qualitative} relationship: we do not have sufficient data to construct a specific quantitative model of the localized increase in radiance for the specific flare considered in this paper. 

\section{Resonant Excitation Methodology}
\label{appendix:calculations}

To construct our emission model, we began with a quiet-sun irradiance from EVE to estimate the photo-pumping of the Fe~\textsc{x} line. From the irradiance measured at Earth we obtained an averaged radiance of 685~$\mathrm{ergs \; cm^{-2} \; s^{-1} \; sr^{-1}}$. At 1.2~R$_\sun$ (measured from Sun center) the disk has a solid angle of 2.81~sr, which results in 1925~$\mathrm{ergs \; cm^{-2} \; s^{-1}}$ photo-pumping radiation. Assuming a disk Gaussian profile of 0.036~\AA\ FWHM and a coronal thermal (isotropic) distribution of ion velocities at 1~MK and no outflows, we obtain from the double integral, for the Fe~\textsc{x} 174.5~\AA\ line, a resonant photo excitation (RPE) rate of 0.14~s$^{-1}$. 

At 1~MK, assuming that electrons and ions have the same
temperature, the collisional excitation (CE)
rate coefficient due to electron impact for the same line 
is $6 \times 10^{-9}$ cm$^{3}$ s$^{-1}$. For an electron density of $10^8$~cm$^{-3}$ the RPE rate is therefore 23\% the CE rate, hence it increases the intensity of the line by that amount. Clearly, if the density were lower, the relative contribution of the RPE would increase.

The local temperature also plays a role as it enters both the calculation of the RPE rate and the CE rate coefficient. However, if, for example, the temperature was 1.5~MK, the contribution of the RPE would be 19\%. What is more important is the distribution of densities along the line of sight, for which we just rely on our MHD model. If the real densities along the line of sight were much lower, the RPE from the solar disk and from the flare would be much higher. As a side note, the effective increase due to RPE by the solar disk is probably underestimated in our calculation, as the Sun was active, and the average radiance could have been up to 30--50\% higher than our baseline estimate.

If we consider now the flare contribution, our assumed
5\% increase in the irradiance is equivalent to an increase of 34~$\mathrm{ergs \; cm^{-2} \; s^{-1}}$ in the averaged disk emission; considering our estimated radius of the flare region, we have  a ratio of the areas of 1890, which yields 64272~$\mathrm{ergs \; cm^{-2} \; s^{-1} \; sr^{-1}}$ due to the flaring region alone. When we approximate this region as an isolated disk with a radius 0.023~R$_\sun$, it has a solid angle of 0.051~sr at our estimated distance of 0.18 R$_\sun$, this results in 3278~$\mathrm{ergs \; cm^{-2} \; s^{-1}}$ RPE radiation from the flare site at the location, i.e., a substantial increase of resonant photoexcitation -- by 70\% over the averaged disk. If we were only considering the plane of sky, instead of the integrated line of sight, this would be the predicted increase in brightness during the peak of the flare.

\bibliography{references}{}

\begin{thebibliography}{}
\expandafter\ifx\csname natexlab\endcsname\relax\def\natexlab#1{#1}\fi
\providecommand{\url}[1]{\href{#1}{#1}}
\providecommand{\dodoi}[1]{doi:~\href{http://doi.org/#1}{\nolinkurl{#1}}}
\providecommand{\doeprint}[1]{\href{http://ascl.net/#1}{\nolinkurl{http://ascl.net/#1}}}
\providecommand{\doarXiv}[1]{\href{https://arxiv.org/abs/#1}{\nolinkurl{https://arxiv.org/abs/#1}}}

\bibitem[{{Auch{\`e}re} {et~al.}(2023){Auch{\`e}re}, {Berghmans}, {Dumesnil},
  {Halain}, {Mercier}, {Rochus}, {Delmotte}, {Fran{\c{c}}ois}, {Hermans},
  {Hervier}, {Kraaikamp}, {Meltchakov}, {Morinaud}, {Philippon}, {Smith},
  {Stegen}, {Verbeeck}, {Zhang}, {Andretta}, {Abbo}, {Buchlin}, {Frassati},
  {Gissot}, {Gyo}, {Harra}, {Jerse}, {Landini}, {Mierla}, {Nicula}, {Parenti},
  {Renotte}, {Romoli}, {Russano}, {Sasso}, {Sch{\"u}hle}, {Schmutz},
  {Soubri{\'e}}, {Susino}, {Teriaca}, {West}, \& {Zhukov}}]{Auchere2023}
{Auch{\`e}re}, F., {Berghmans}, D., {Dumesnil}, C., {et~al.} 2023, \aap, 674,
  A127, \dodoi{10.1051/0004-6361/202346039}

\bibitem[{{Boe} {et~al.}(2023){Boe}, {Downs}, \& {Habbal}}]{Boe2023}
{Boe}, B., {Downs}, C., \& {Habbal}, S. 2023, \apj, 951, 55,
  \dodoi{10.3847/1538-4357/acd10b}

\bibitem[{{Boe} {et~al.}(2021){Boe}, {Habbal}, {Downs}, \&
  {Druckm{\"u}ller}}]{boe21}
{Boe}, B., {Habbal}, S., {Downs}, C., \& {Druckm{\"u}ller}, M. 2021, \apj, 912,
  44, \dodoi{10.3847/1538-4357/abea79}

\bibitem[{{Boe} {et~al.}(2022){Boe}, {Habbal}, {Downs}, \&
  {Druckm{\"u}ller}}]{boe22}
---. 2022, \apj, 935, 173, \dodoi{10.3847/1538-4357/ac8101}

\bibitem[{{Bonte}(2014)}]{Bonte2013}
{Bonte}, K. 2014, PhD thesis, Katholieke University of Leuven, Belgium

\bibitem[{{Brueckner} {et~al.}(1995){Brueckner}, {Howard}, {Koomen},
  {Korendyke}, {Michels}, {Moses}, {Socker}, {Dere}, {Lamy}, {Llebaria},
  {Bout}, {Schwenn}, {Simnett}, {Bedford}, \& {Eyles}}]{Brueckner1995}
{Brueckner}, G.~E., {Howard}, R.~A., {Koomen}, M.~J., {et~al.} 1995, \solphys,
  162, 357, \dodoi{10.1007/BF00733434}

\bibitem[{{Cheung} {et~al.}(2015){Cheung}, {Boerner}, {Schrijver}, {Testa},
  {Chen}, {Peter}, \& {Malanushenko}}]{Cheung2015}
{Cheung}, M. C.~M., {Boerner}, P., {Schrijver}, C.~J., {et~al.} 2015, \apj,
  807, 143, \dodoi{10.1088/0004-637X/807/2/143}

\bibitem[{{Darnel} {et~al.}(2022){Darnel}, {Seaton}, {Bethge}, {Rachmeler},
  {Jarvis}, {Hill}, {Peck}, {Hughes}, {Shapiro}, {Riley}, {Vasudevan}, {Shing},
  {Koener}, {Edwards}, {Mathur}, \& {Timothy}}]{Darnel2022}
{Darnel}, J.~M., {Seaton}, D.~B., {Bethge}, C., {et~al.} 2022, Space Weather,
  in prep.

\bibitem[{{DeForest} {et~al.}(2009){DeForest}, {Martens}, \&
  {Wills-Davey}}]{deforest2009}
{DeForest}, C.~E., {Martens}, P.~C.~H., \& {Wills-Davey}, M.~J. 2009, \apj,
  690, 1264, \dodoi{10.1088/0004-637X/690/2/1264}

\bibitem[{{Del Zanna}(2019)}]{delzanna19}
{Del Zanna}, G. 2019, \aap, 624, A36, \dodoi{10.1051/0004-6361/201834842}

\bibitem[{{Del Zanna} {et~al.}(2021){Del Zanna}, {Dere}, {Young}, \&
  {Landi}}]{delzanna21_chianti10}
{Del Zanna}, G., {Dere}, K.~P., {Young}, P.~R., \& {Landi}, E. 2021, \apj, 909,
  38, \dodoi{10.3847/1538-4357/abd8ce}

\bibitem[{{Del Zanna} \& {Mason}(2018)}]{delzanna2018a}
{Del Zanna}, G., \& {Mason}, H.~E. 2018, Living Reviews in Solar Physics, 15,
  5, \dodoi{10.1007/s41116-018-0015-3}

\bibitem[{{Del Zanna} {et~al.}(2018){Del Zanna}, {Raymond}, {Andretta},
  {Telloni}, \& {Golub}}]{delzanna2018b}
{Del Zanna}, G., {Raymond}, J., {Andretta}, V., {Telloni}, D., \& {Golub}, L.
  2018, \apj, 865, 132, \dodoi{10.3847/1538-4357/aadcf1}

\bibitem[{{Del Zanna} \& {Woods}(2013)}]{DelZanna_Woods_2013}
{Del Zanna}, G., \& {Woods}, T.~N. 2013, \aap, 555, A59,
  \dodoi{10.1051/0004-6361/201220988}

\bibitem[{{Dere} {et~al.}(1997){Dere}, {Landi}, {Mason}, {Monsignori Fossi}, \&
  {Young}}]{dere97}
{Dere}, K.~P., {Landi}, E., {Mason}, H.~E., {Monsignori Fossi}, B.~C., \&
  {Young}, P.~R. 1997, \aaps, 125, 149, \dodoi{10.1051/aas:1997368}

\bibitem[{{Dolei} {et~al.}(2019){Dolei}, {Spadaro}, {Ventura}, {Bemporad},
  {Andretta}, {Sasso}, {Susino}, {Antonucci}, {Da Deppo}, {Fineschi},
  {Frassetto}, {Landini}, {Naletto}, {Nicolini}, {Pancrazzi}, \&
  {Romoli}}]{dolei19}
{Dolei}, S., {Spadaro}, D., {Ventura}, R., {et~al.} 2019, \aap, 627, A18,
  \dodoi{10.1051/0004-6361/201935048}

\bibitem[{{Downs} {et~al.}(2016){Downs}, {Lionello}, {Miki{\'c}}, {Linker}, \&
  {Velli}}]{downs16}
{Downs}, C., {Lionello}, R., {Miki{\'c}}, Z., {Linker}, J.~A., \& {Velli}, M.
  2016, \apj, 832, 180, \dodoi{10.3847/0004-637X/832/2/180}

\bibitem[{{Eparvier} {et~al.}(2015){Eparvier}, {Chamberlin}, {Woods}, \&
  {Thiemann}}]{Eparvier2015}
{Eparvier}, F.~G., {Chamberlin}, P.~C., {Woods}, T.~N., \& {Thiemann}, E.~M.~B.
  2015, \ssr, 195, 293, \dodoi{10.1007/s11214-015-0195-2}

\bibitem[{{Gilly} \& {Cranmer}(2020)}]{Gilly2020}
{Gilly}, C.~R., \& {Cranmer}, S.~R. 2020, \apj, 901, 150,
  \dodoi{10.3847/1538-4357/abb1ad}

\bibitem[{{Golub} {et~al.}(2007){Golub}, {Deluca}, {Austin}, {Bookbinder},
  {Caldwell}, {Cheimets}, {Cirtain}, {Cosmo}, {Reid}, {Sette}, {Weber},
  {Sakao}, {Kano}, {Shibasaki}, {Hara}, {Tsuneta}, {Kumagai}, {Tamura},
  {Shimojo}, {McCracken}, {Carpenter}, {Haight}, {Siler}, {Wright}, {Tucker},
  {Rutledge}, {Barbera}, {Peres}, \& {Varisco}}]{Golub2007}
{Golub}, L., {Deluca}, E., {Austin}, G., {et~al.} 2007, \solphys, 243, 63,
  \dodoi{10.1007/s11207-007-0182-1}

\bibitem[{{Goryaev} {et~al.}(2014){Goryaev}, {Slemzin}, {Vainshtein}, \&
  {Williams}}]{goryaev2014}
{Goryaev}, F., {Slemzin}, V., {Vainshtein}, L., \& {Williams}, D.~R. 2014,
  \apj, 781, 100, \dodoi{10.1088/0004-637X/781/2/100}

\bibitem[{{Goryaev} {et~al.}(2018){Goryaev}, {Slemzin}, {Rodkin}, {D'Huys},
  {Podladchikova}, \& {West}}]{goryaev2018}
{Goryaev}, F.~F., {Slemzin}, V.~A., {Rodkin}, D.~G., {et~al.} 2018, \apjl, 856,
  L38, \dodoi{10.3847/2041-8213/aab849}

\bibitem[{{Halain} {et~al.}(2013){Halain}, {Berghmans}, {Seaton}, {Nicula}, {De
  Groof}, {Mierla}, {Mazzoli}, {Defise}, \& {Rochus}}]{Halain2013}
{Halain}, J.~P., {Berghmans}, D., {Seaton}, D.~B., {et~al.} 2013, \solphys,
  286, 67, \dodoi{10.1007/s11207-012-0183-6}

\bibitem[{{Handy} {et~al.}(1999){Handy}, {Acton}, {Kankelborg}, {Wolfson},
  {Akin}, {Bruner}, {Caravalho}, {Catura}, {Chevalier}, {Duncan}, {Edwards},
  {Feinstein}, {Freeland}, {Friedlaender}, {Hoffmann}, {Hurlburt}, {Jurcevich},
  {Katz}, {Kelly}, {Lemen}, {Levay}, {Lindgren}, {Mathur}, {Meyer}, {Morrison},
  {Morrison}, {Nightingale}, {Pope}, {Rehse}, {Schrijver}, {Shine}, {Shing},
  {Strong}, {Tarbell}, {Title}, {Torgerson}, {Golub}, {Bookbinder}, {Caldwell},
  {Cheimets}, {Davis}, {Deluca}, {McMullen}, {Warren}, {Amato}, {Fisher},
  {Maldonado}, \& {Parkinson}}]{handy1999}
{Handy}, B.~N., {Acton}, L.~W., {Kankelborg}, C.~C., {et~al.} 1999, \solphys,
  187, 229, \dodoi{10.1023/A:1005166902804}

\bibitem[{{Hock}(2012)}]{hock12_thesis}
{Hock}, R.~A. 2012, PhD thesis, University of Colorado, Boulder

\bibitem[{{Howard} {et~al.}(2008){Howard}, {Moses}, {Vourlidas}, {Newmark},
  {Socker}, {Plunkett}, {Korendyke}, {Cook}, {Hurley}, {Davila}, {Thompson},
  {St Cyr}, {Mentzell}, {Mehalick}, {Lemen}, {Wuelser}, {Duncan}, {Tarbell},
  {Wolfson}, {Moore}, {Harrison}, {Waltham}, {Lang}, {Davis}, {Eyles},
  {Mapson-Menard}, {Simnett}, {Halain}, {Defise}, {Mazy}, {Rochus}, {Mercier},
  {Ravet}, {Delmotte}, {Auchere}, {Delaboudiniere}, {Bothmer}, {Deutsch},
  {Wang}, {Rich}, {Cooper}, {Stephens}, {Maahs}, {Baugh}, {McMullin}, \&
  {Carter}}]{Howard2008}
{Howard}, R.~A., {Moses}, J.~D., {Vourlidas}, A., {et~al.} 2008, \ssr, 136, 67,
  \dodoi{10.1007/s11214-008-9341-4}

\bibitem[{{Judge}(1998)}]{Judge1998}
{Judge}, P.~G. 1998, \apj, 500, 1009, \dodoi{10.1086/305775}

\bibitem[{{Lionello} {et~al.}(2009){Lionello}, {Linker}, \&
  {Miki{\'c}}}]{lionello09}
{Lionello}, R., {Linker}, J.~A., \& {Miki{\'c}}, Z. 2009, \apj, 690, 902,
  \dodoi{10.1088/0004-637X/690/1/902}

\bibitem[{{Mart{\'\i}nez-Galarce} {et~al.}(2010){Mart{\'\i}nez-Galarce},
  {Harvey}, {Bruner}, {Lemen}, {Gullikson}, {Soufli}, {Prast}, \&
  {Khatri}}]{martinez-galarce2010}
{Mart{\'\i}nez-Galarce}, D., {Harvey}, J., {Bruner}, M., {et~al.} 2010, in
  Society of Photo-Optical Instrumentation Engineers (SPIE) Conference Series,
  Vol. 7732, Space Telescopes and Instrumentation 2010: Ultraviolet to Gamma
  Ray, ed. M.~{Arnaud}, S.~S. {Murray}, \& T.~{Takahashi}, 773237,
  \dodoi{10.1117/12.864577}

\bibitem[{{Mason} {et~al.}(2019){Mason}, {Attie}, {Arge}, {Thompson}, \&
  {Woods}}]{mason19_catalog}
{Mason}, J.~P., {Attie}, R., {Arge}, C.~N., {Thompson}, B., \& {Woods}, T.~N.
  2019, \apjs, 244, 13, \dodoi{10.3847/1538-4365/ab380e}

\bibitem[{{Mason} {et~al.}(2014){Mason}, {Woods}, {Caspi}, {Thompson}, \&
  {Hock}}]{Mason2014}
{Mason}, J.~P., {Woods}, T.~N., {Caspi}, A., {Thompson}, B.~J., \& {Hock},
  R.~A. 2014, \apj, 789, 61, \dodoi{10.1088/0004-637X/789/1/61}

\bibitem[{{Mason} {et~al.}(2021){Mason}, {Chamberlin}, {Seaton}, {Burkepile},
  {Colaninno}, {Dissauer}, {Eparvier}, {Fan}, {Gibson}, {Jones}, {Kay}, {Kirk},
  {Kohnert}, {Pesnell}, {Thompson}, {Veronig}, {West}, {Windt}, \&
  {Woods}}]{Mason2021}
{Mason}, J.~P., {Chamberlin}, P.~C., {Seaton}, D., {et~al.} 2021, Journal of
  Space Weather and Space Climate, 11, 20, \dodoi{10.1051/swsc/2021004}

\bibitem[{{Mierla} {et~al.}(2022){Mierla}, {Zhukov}, {Berghmans}, {Parenti},
  {Auch{\`e}re}, {Heinzel}, {Seaton}, {Palmerio}, {Jej{\v{c}}i{\v{c}}},
  {Janssens}, {Kraaikamp}, {Nicula}, {Long}, {Hayes}, {Jebaraj}, {Talpeanu},
  {D'Huys}, {Dolla}, {Gissot}, {Magdaleni{\'c}}, {Rodriguez}, {Shestov},
  {Stegen}, {Verbeeck}, {Sasso}, {Romoli}, \& {Andretta}}]{Mierla2022}
{Mierla}, M., {Zhukov}, A.~N., {Berghmans}, D., {et~al.} 2022, \aap, 662, L5,
  \dodoi{10.1051/0004-6361/202244020}

\bibitem[{{Miki{\'c}} {et~al.}(1999){Miki{\'c}}, {Linker}, {Schnack},
  {Lionello}, \& {Tarditi}}]{mikic99}
{Miki{\'c}}, Z., {Linker}, J.~A., {Schnack}, D.~D., {Lionello}, R., \&
  {Tarditi}, A. 1999, Physics of Plasmas, 6, 2217, \dodoi{10.1063/1.873474}

\bibitem[{{Miki{\'c}} {et~al.}(2018){Miki{\'c}}, {Downs}, {Linker}, {Caplan},
  {Mackay}, {Upton}, {Riley}, {Lionello}, {T{\"o}r{\"o}k}, {Titov}, {Wijaya},
  {Druckm{\"u}ller}, {Pasachoff}, \& {Carlos}}]{mikic18}
{Miki{\'c}}, Z., {Downs}, C., {Linker}, J.~A., {et~al.} 2018, Nature Astronomy,
  2, 913, \dodoi{10.1038/s41550-018-0562-5}

\bibitem[{{Milligan} \& {McElroy}(2013)}]{Milligan2013}
{Milligan}, R.~O., \& {McElroy}, S.~A. 2013, \apj, 777, 12,
  \dodoi{10.1088/0004-637X/777/1/12}

\bibitem[{{O'Dwyer} {et~al.}(2010){O'Dwyer}, {Del Zanna}, {Mason}, {Weber}, \&
  {Tripathi}}]{ODwyer2010}
{O'Dwyer}, B., {Del Zanna}, G., {Mason}, H.~E., {Weber}, M.~A., \& {Tripathi},
  D. 2010, \aap, 521, A21, \dodoi{10.1051/0004-6361/201014872}

\bibitem[{{Plowman} \& {Caspi}(2020)}]{Plowman2020}
{Plowman}, J., \& {Caspi}, A. 2020, \apj, 905, 17,
  \dodoi{10.3847/1538-4357/abc260}

\bibitem[{{Raftery} {et~al.}(2013){Raftery}, {Bloomfield}, {Gallagher},
  {Seaton}, {Berghmans}, \& {De Groof}}]{Raftery2013}
{Raftery}, C.~L., {Bloomfield}, D.~S., {Gallagher}, P.~T., {et~al.} 2013,
  \solphys, 286, 111, \dodoi{10.1007/s11207-013-0266-z}

\bibitem[{{Robbrecht} {et~al.}(2009){Robbrecht}, {Berghmans}, \& {Van der
  Linden}}]{Robbrecht2009}
{Robbrecht}, E., {Berghmans}, D., \& {Van der Linden}, R.~A.~M. 2009, \apj,
  691, 1222, \dodoi{10.1088/0004-637X/691/2/1222}

\bibitem[{{Rochus} {et~al.}(2020){Rochus}, {Auch{\`e}re}, {Berghmans}, {Harra},
  {Schmutz}, {Sch{\"u}hle}, {Addison}, {Appourchaux}, {Aznar Cuadrado},
  {Baker}, {Barbay}, {Bates}, {BenMoussa}, {Bergmann}, {Beurthe}, {Borgo},
  {Bonte}, {Bouzit}, {Bradley}, {B{\"u}chel}, {Buchlin}, {B{\"u}chner},
  {Cab{\'e}}, {Cadiergues}, {Chaigneau}, {Chares}, {Choque Cortez}, {Coker},
  {Condamin}, {Coumar}, {Curdt}, {Cutler}, {Davies}, {Davison}, {Defise}, {Del
  Zanna}, {Delmotte}, {Delouille}, {Dolla}, {Dumesnil}, {D{\"u}rig}, {Enge},
  {Fran{\c{c}}ois}, {Fourmond}, {Gillis}, {Giordanengo}, {Gissot}, {Green},
  {Guerreiro}, {Guilbaud}, {Gyo}, {Haberreiter}, {Hafiz}, {Hailey}, {Halain},
  {Hansotte}, {Hecquet}, {Heerlein}, {Hellin}, {Hemsley}, {Hermans}, {Hervier},
  {Hochedez}, {Houbrechts}, {Ihsan}, {Jacques}, {J{\'e}r{\^o}me}, {Jones},
  {Kahle}, {Kennedy}, {Klaproth}, {Kolleck}, {Koller}, {Kotsialos},
  {Kraaikamp}, {Langer}, {Lawrenson}, {Le Clech'}, {Lenaerts}, {Liebecq},
  {Linder}, {Long}, {Mampaey}, {Markiewicz-Innes}, {Marquet}, {Marsch},
  {Matthews}, {Mazy}, {Mazzoli}, {Meining}, {Meltchakov}, {Mercier}, {Meyer},
  {Monecke}, {Monfort}, {Morinaud}, {Moron}, {Mountney}, {M{\"u}ller},
  {Nicula}, {Parenti}, {Peter}, {Pfiffner}, {Philippon}, {Phillips},
  {Plesseria}, {Pylyser}, {Rabecki}, {Ravet-Krill}, {Rebellato}, {Renotte},
  {Rodriguez}, {Roose}, {Rosin}, {Rossi}, {Roth}, {Rouesnel}, {Roulliay},
  {Rousseau}, {Ruane}, {Scanlan}, {Schlatter}, {Seaton}, {Silliman}, {Smit},
  {Smith}, {Solanki}, {Spescha}, {Spencer}, {Stegen}, {Stockman}, {Szwec},
  {Tamiatto}, {Tandy}, {Teriaca}, {Theobald}, {Tychon}, {van Driel-Gesztelyi},
  {Verbeeck}, {Vial}, {Werner}, {West}, {Westwood}, {Wiegelmann}, {Willis},
  {Winter}, {Zerr}, {Zhang}, \& {Zhukov}}]{Rochus2020}
{Rochus}, P., {Auch{\`e}re}, F., {Berghmans}, D., {et~al.} 2020, \aap, 642, A8,
  \dodoi{10.1051/0004-6361/201936663}

\bibitem[{{Savage} {et~al.}(2012){Savage}, {McKenzie}, \&
  {Reeves}}]{Savage2012}
{Savage}, S.~L., {McKenzie}, D.~E., \& {Reeves}, K.~K. 2012, \apjl, 747, L40,
  \dodoi{10.1088/2041-8205/747/2/L40}

\bibitem[{{Schrijver} \& {McMullen}(2000)}]{schrijver_mcmullen_2000}
{Schrijver}, C.~J., \& {McMullen}, R.~A. 2000, \apj, 531, 1121,
  \dodoi{10.1086/308497}

\bibitem[{{Schwartz} {et~al.}(2014){Schwartz}, {Torre}, \&
  {Piana}}]{schwartz2014}
{Schwartz}, R.~A., {Torre}, G., \& {Piana}, M. 2014, \apjl, 793, L23,
  \dodoi{10.1088/2041-8205/793/2/L23}

\bibitem[{{Seaton} {et~al.}(2013{\natexlab{a}}){Seaton}, {De Groof}, {Shearer},
  {Berghmans}, \& {Nicula}}]{seaton2013}
{Seaton}, D.~B., {De Groof}, A., {Shearer}, P., {Berghmans}, D., \& {Nicula},
  B. 2013{\natexlab{a}}, \apj, 777, 72, \dodoi{10.1088/0004-637X/777/1/72}

\bibitem[{{Seaton} \& {Forbes}(2009)}]{Seaton2009}
{Seaton}, D.~B., \& {Forbes}, T.~G. 2009, \apj, 701, 348,
  \dodoi{10.1088/0004-637X/701/1/348}

\bibitem[{{Seaton} {et~al.}(2013{\natexlab{b}}){Seaton}, {Berghmans}, {Nicula},
  {Halain}, {De Groof}, {Thibert}, {Bloomfield}, {Raftery}, {Gallagher},
  {Auch{\`e}re}, {Defise}, {D'Huys}, {Lecat}, {Mazy}, {Rochus}, {Rossi},
  {Sch{\"u}hle}, {Slemzin}, {Yalim}, \& {Zender}}]{SeatonSwap2013}
{Seaton}, D.~B., {Berghmans}, D., {Nicula}, B., {et~al.} 2013{\natexlab{b}},
  \solphys, 286, 43, \dodoi{10.1007/s11207-012-0114-6}

\bibitem[{{Seaton} {et~al.}(2021){Seaton}, {Hughes}, {Tadikonda}, {Caspi},
  {DeForest}, {Krimchansky}, {Hurlburt}, {Seguin}, \& {Slater}}]{Seaton2021}
{Seaton}, D.~B., {Hughes}, J.~M., {Tadikonda}, S.~K., {et~al.} 2021, Nature
  Astronomy, \dodoi{10.1038/s41550-021-01427-8}

\bibitem[{{Shen} {et~al.}(2022){Shen}, {Chen}, {Reeves}, {Yu}, {Polito}, \&
  {Xie}}]{Shen2022}
{Shen}, C., {Chen}, B., {Reeves}, K.~K., {et~al.} 2022, Nature Astronomy, 6,
  317, \dodoi{10.1038/s41550-021-01570-2}

\bibitem[{{Strachan} {et~al.}(1993){Strachan}, {Kohl}, {Weiser}, {Withbroe}, \&
  {Munro}}]{Strachan1993}
{Strachan}, L., {Kohl}, J.~L., {Weiser}, H., {Withbroe}, G.~L., \& {Munro},
  R.~H. 1993, \apj, 412, 410, \dodoi{10.1086/172930}

\bibitem[{{Thompson}(2009)}]{Thompson2009}
{Thompson}, W.~T. 2009, \icarus, 200, 351, \dodoi{10.1016/j.icarus.2008.12.011}

\bibitem[{{West} {et~al.}(2023){West}, {Seaton}, {Wexler}, {Raymond}, {Del
  Zanna}, {Rivera}, {Kobelski}, {Chen}, {DeForest}, {Golub}, {Caspi}, {Gilly},
  {Kooi}, {Meyer}, {Alterman}, {Alzate}, {Andretta}, {Auch{\`e}re}, {Banerjee},
  {Berghmans}, {Chamberlin}, {Chitta}, {Downs}, {Giordano}, {Harra},
  {Higginson}, {Howard}, {Kumar}, {Mason}, {Mason}, {Morton}, {Nykyri},
  {Patel}, {Rachmeler}, {Reardon}, {Reeves}, {Savage}, {Thompson}, {Van
  Kooten}, {Viall}, {Vourlidas}, \& {Zhukov}}]{West2023}
{West}, M.~J., {Seaton}, D.~B., {Wexler}, D.~B., {et~al.} 2023, \solphys, 298,
  78, \dodoi{10.1007/s11207-023-02170-1}

\bibitem[{Woods {et~al.}(2010)Woods, Eparvier, Hock, Jones, Woodraska, Judge,
  Didkovsky, Lean, Mariska, Warren, McMullin, Chamberlin, Berthiaume, Bailey,
  Fuller-Rowell, Sojka, Tobiska, \& Viereck}]{Woods2010}
Woods, T.~N., Eparvier, F.~G., Hock, R.~A., {et~al.} 2010, Solar Physics, 275,
  115, \dodoi{10.1007/s11207-009-9487-6}

\bibitem[{{Woods} {et~al.}(2012){Woods}, {Eparvier}, {Hock}, {Jones},
  {Woodraska}, {Judge}, {Didkovsky}, {Lean}, {Mariska}, {Warren}, {McMullin},
  {Chamberlin}, {Berthiaume}, {Bailey}, {Fuller-Rowell}, {Sojka}, {Tobiska}, \&
  {Viereck}}]{woods12}
{Woods}, T.~N., {Eparvier}, F.~G., {Hock}, R., {et~al.} 2012, \solphys, 275,
  115, \dodoi{10.1007/s11207-009-9487-6}

\bibitem[{{Yokoyama} \& {Shibata}(2001)}]{Yokoyama2001}
{Yokoyama}, T., \& {Shibata}, K. 2001, \apj, 549, 1160, \dodoi{10.1086/319440}

\end{thebibliography}
\bibliographystyle{aasjournal}



\end{document}